\documentclass[12pt,preprint]{aastex}

\newcommand{\ea}{{\it et al.}}

\newcommand{\jcomp}{\rmfamily{J. Comp. Phys.}}

\newcommand{\kms}{km~$\rm{s}^{-1}$}

\newcommand{\beq}{\begin{equation}}
\newcommand{\eeq}{\end{equation}}
\newcommand{\bdm}{\begin{displaymath}}
\newcommand{\edm}{\end{displaymath}}

\slugcomment{Submitted to the Astrophysical Journal}

\begin{document}

\title{Hydrodynamic Interaction of Strong Shocks with Inhomogeneous
Media - I: Adiabatic Case}

\author{A.Y. Poludnenko\altaffilmark{1}, A. Frank\altaffilmark{2},
E.G. Blackman
\altaffilmark{3}}
\affil{Department of Physics and Astronomy,\\
       University of Rochester, Rochester, NY 14627-0171}
\altaffiltext{1}{wma@pas.rochester.edu}
\altaffiltext{2}{afrank@pas.rochester.edu}
\altaffiltext{3}{blackman@pas.rochester.edu}

\begin{abstract}
Many astrophysical flows occur in inhomogeneous (clumpy) media. We
present results of a numerical study of steady, planar shocks
interacting with a system of embedded cylindrical clouds.  Our
study uses a two-dimensional geometry. Our numerical code uses an
adaptive mesh refinement allowing us to achieve sufficiently high
resolution both at the largest and the smallest scales. We neglect
any radiative losses, heat conduction, and gravitational forces.
Detailed analysis of the simulations shows that interaction of
embedded inhomogeneities with the shock/postshock wind depends
primarily on the thickness of the cloud layer and arrangement of
the clouds in the layer. The total cloud mass and the total number
of individual clouds is not a significant factor. We define two
classes of cloud distributions: thin and thick layers. We define
the critical cloud separation along the direction of the flow and
perpendicular to it distinguishing between the interacting and
noninteracting regimes of cloud evolution. Finally we discuss
mass-loading and mixing in such systems.

\end{abstract}


\section{INTRODUCTION}

Mass outflows play a critical role in many astrophysical systems
ranging from stars to the most distant active galaxies. Virtually
all studies of mass outflows to date have focused on flows
in homogeneous media. However, the typical astrophysical medium is
inhomogeneous with the "clumps" or "clouds" arising on a variety
of scales.  These inhomogeneities may arise due to initial
fluctuations of the ambient mass distribution, the action of
instabilities, variations in the flow source, etc. Whatever the
origin of the clumps their effect can be dramatic. The presence of
inhomogeneities can introduce not only quantitative but also
qualitative changes to the overall dynamics of the flow.

A number of studies have attempted to understand the role of
embedded inhomogeneities via (primarily) analytical methods
\citep{HD86}, \citep{HD88}, \citep{DH92}, \citep{DH94}. In these
pioneering works it was suggested that interactions of the global
flow with inhomogeneities may cause significant changes in the
physical, dynamical, and even chemical state of the system. Two
major consequences of the presence of clumps are mass-loading
(i.e. seeding of material, ablated from the surface of
inhomogeneities, into the global flow) and transition of the global
flow into a transonic regime irrespective of the initial
conditions. The papers cited above considered the potential
effects of mass-loading on the global properties of a number
objects in which inhomogeneities can be resolved. Such objects
include planetary nebulae, e.g. NGC 2392 \citep{ODell90},
\citep{Phillips99}, and NGC 7293 \citep{ODell98}, and Wolf-Rayet
stars, and primarily RCW58, which is believed to be mass-loading
dominated \citep{HD86}.

A number of numerical studies of {\it single clump} interactions have
been performed (\citep{KMC} (hereafter KMC), \citep{Ander94},
\citep{Jones96}, \citep{Gregori99}, \citep{Gregori2000}, \citep{Jun99}, 
\citep{Miniati99}, \citep{Lim99}). In these papers the basic 
hydrodynamics or MHD of wind-clump and shock-clump physics have been
detailed (often with microphysical processes included).  A few studies
of shock waves overrunning over multiple clumps exist in the
literature (e.g. \citep{Jun96}). A detailed study of multiple clumps
however, where an attempt is made to articulate basic physical
processes and differentiate various parameter regimes, has not yet
been carried out. In this paper, (and those which follow), we address
the problem of clumpy flows providing a description of the dynamics of
multiple dense clouds interacting with a strong, steady, planar
shock. 

The large parameter space and complexity of the problem require
significant computational effort. To provide the necessary resolution
of the flow we have used an adaptive mesh refinement method. This is a
relatively new computational technology and because of this we have
chosen to investigate so-called adiabatic flows in which radiative
cooling is not considered.  In this regard our approach is similar to
that described by \cite{KMC} for single clumps and we will utilize
their results in understanding our multi-clump simulations. We note
that preliminary results, appropriate to AGN, were presented in
\citep{PolAGN01}.

The plan of the paper is as follows.  In Section 2 we describe the
numerical experiments, the code used, and the formulation of the
problem. In Section 3.1 we consider the general properties of the
shock-cloud interaction in the context of the multi-cloud systems,
primarily we focus on the four major phases of the interaction
process. In Section 3.2 we discuss the role of cloud distribution
in determining the dynamics of the system evolution. In Section
3.3 we define several key parameters, that allow us to distinguish
between various regimes of shock-cloud interaction. Finally, in
Section 3.4 we address the issue of mass-loading in such systems.

\section{NUMERICAL EXPERIMENTS}

\subsection{Description of the Code Used}

The code we used for this project is the AMRCLAW package which
implements an adaptive mesh refinement algorithm for the equations
of gas dynamics (\citep{Berger97}, \citep{Berger85},
\citep{Berger89}, \citep{Berger94}).  In the AMRCLAW approach, the
computational domain is associated with a logically rectangular
grid that represents the lowest level of refinement (level 1) and
that embeds the nested sequence of logically rectangular meshes
with finer resolution (levels 2,3,...). The temporal and spatial
steps of all grids at a level L are refined with respect to the
level L-1 grids by the same factor, typically 4 in our
calculations. The mesh ratios $\Delta t/ \Delta x$ and
$\Delta t/ \Delta y$ are then the same on all grids, ensuring
stability with explicit difference schemes.

The core of the code - the AMR module - scans each refinement
level every \emph{k} time steps and regenerates all nested higher
level grids in order to track the moving features of the flow. Two
criteria are used to define cells requiring refinement: Richardson
extrapolation and steepest gradient. The first criterion ensures
that the local truncation error does not exceed some predefined
tolerance. This is done via comparison of the solution obtained by
taking 2 time steps on the existing grid with one computed by
taking 1 time step on a grid that is twice as coarse in each
direction. The second criterion ensures that the maximum of the
gradients of all state variables does not exceed some predefined
value and guarantees that sufficient refinement is achieved in
such regions of the flow as shock waves, boundary layers, etc.
Flagged cells are then organized into rectangular grid patches in
a manner that provides a reasonable compromise between the size
and the total number of individual patches. Finally, the AMR
module of the code ensures that global conservation is preserved
at grid interfaces via introduction of a conservative flux
correction.

After the grid hierarchy is formed, each grid is forwarded to the
integration module. This module considers every grid as an independent
physical domain. The boundary conditions are obtained either from the
physical boundary conditions of the computational domain or via
interpolation from the neighbouring cells of the next lowest refinement
level, depending on the grid location. This approach allows us to
separate logically the AMR and integration modules, which facilitates
incorporation of new features into the code.  The integration proceeds
by grid level starting at level 1, which is integrated over a time
step, then at level 2 (it should be integrated over
$R_{1}$=$x_{2}$/$x_{1}$= $y_{2}$/$y_{1}$=$t_{2}$/$t_{1}$ time steps to
catch up), and so on. The solution on each grid is advanced via a
second-order accurate Godunov-type finite volume method in which
second-order accuracy is achieved via flux-limiting and proper
consideration of transverse wave propagation. The multi-dimensional
wave propagation algorithm is based on the traditional dimensional
splitting with the Riemann problem solved in each dimension by means
of a Roe-approximate Riemann solver \citep{Leveque97}. It should be
noted, that our implementation of the Riemann solver, based on the Roe
linearization, does not use any additional procedures to ensure
satisfaction of the entropy condition, as usually employed for this
type of Riemann solver. Our analysis shows that the numerical
diffusion present in the system is sufficient to prevent
entropy-violating waves from propagating in the system.

The hydrodynamic equations we solve are appropriate to a single-fluid
system, although a passive tracer is introduced in order to track
advection and mixing of the cloud material. This was implemented as an
additional wave family in the Roe solver.

Our numerical experiments were performed on a coarse grid with the
resolution of 50$\times$100 cells and with the maximum number of
refinement levels equal to 3 (meaning that the coarse grid associated
with the computational domain embeds not more than two nested higher
resolution levels). Each higher level has a temporal and spatial step
refined by the factor of 4 in comparison with the next lowest level
and we kept this refinement ratio constant for all levels. Such setup
provides the \emph{equivalent resolution}\footnote{By equivalent
resolution hereafter we mean the resolution of a uniform grid covering
all of the computational domain and possessing the temporal and
spatial step of the highest refinement level.} of 800$\times$1600
cells. In order to facilitate comparison of our numerical experiments
with those of KMC, we will describe the resolution not in terms of the
equivalent resolution but in terms of the number of cells that fit in
the original maximal cloud radius $a_{0}$, following the convention of
KMC. Then all of the runs described here in our paper have 32 cells
per cloud radius.

KMC suggested that a minimum resolution of 120 cells per cloud radius
is necessary. We have performed the simulations of the cloud-shock
interaction with the resolution of 120, 75, and 55 cells per cloud
radius. Although we will not describe the details of those runs in
this paper, the principal difference between the cases with maximum
and minimum resolution, i.e. 120 and 32 cells per cloud radius, is the
rate of instability formation at the boundary layers\footnote{For the
case of lower resolution the lower rate of instability formation may
be somewhat compensated by the use of the compressive flux
limiters.}. This does not seem to have any significant effect on the
global properties of the interaction or the averaged characteristics
of the individual cloud ablation processes. Therefore, we find the
resolution of 30 cells per cloud radius and above to represent
accurately the global properties of the interaction process under
consideration. Moreover, 30 cells per cloud radius is a reasonable
compromise between maximizing the size of the computational domain and
capturing as many small-scale features of the interaction process as
possible.

Finally, another aspect of this problem is the connection between
the spatial resolution (which naturally sets the smallest scale
resolvable in the simulations) and the diffusion and thermal
conduction length scales. As we will see in section 3.1.4, viscous
diffusion and thermal conduction in a real physical system operate
at length scales comparable to the size of a computational cell at
the highest refinement level used in our simulations. Therefore,
in a real system, features smaller than the ones that can be
resolved with our resolution could not survive over the dynamical
time scales relevant to the problem. We will address this in
greater detail when we discuss the mixing phase of cloud
evolution.

\subsection{Formulation of the Problem}

We set up a two-dimensional computational volume, associated with the
initial condition of $N$ different clouds of radius $a_{i}$ and
density $\rho_{i}$ embedded in the ambient medium of density
$\rho_{a}$, and an incident shock wave. Since all of the experiments
were performed in the Cartesian geometry, the clouds are actually
cross-sections of the infinitely long cylinders. We will address the
importance of the cloud shape in more detail in subsequent work where
we will consider the fully 3-dimensional case of the shock interaction
with spherical clouds. Denoting the maximum cloud radius present in
the system as $a_{max}$, our computational domain is
$25a_{max}\times50a_{max}$.  This allows us to track the dynamical
evolution of the system over greater temporal and spatial intervals
compared to the $6a_{max}\times16a_{max}$ domain, considered by KMC.

\begin{figure}
\plotone{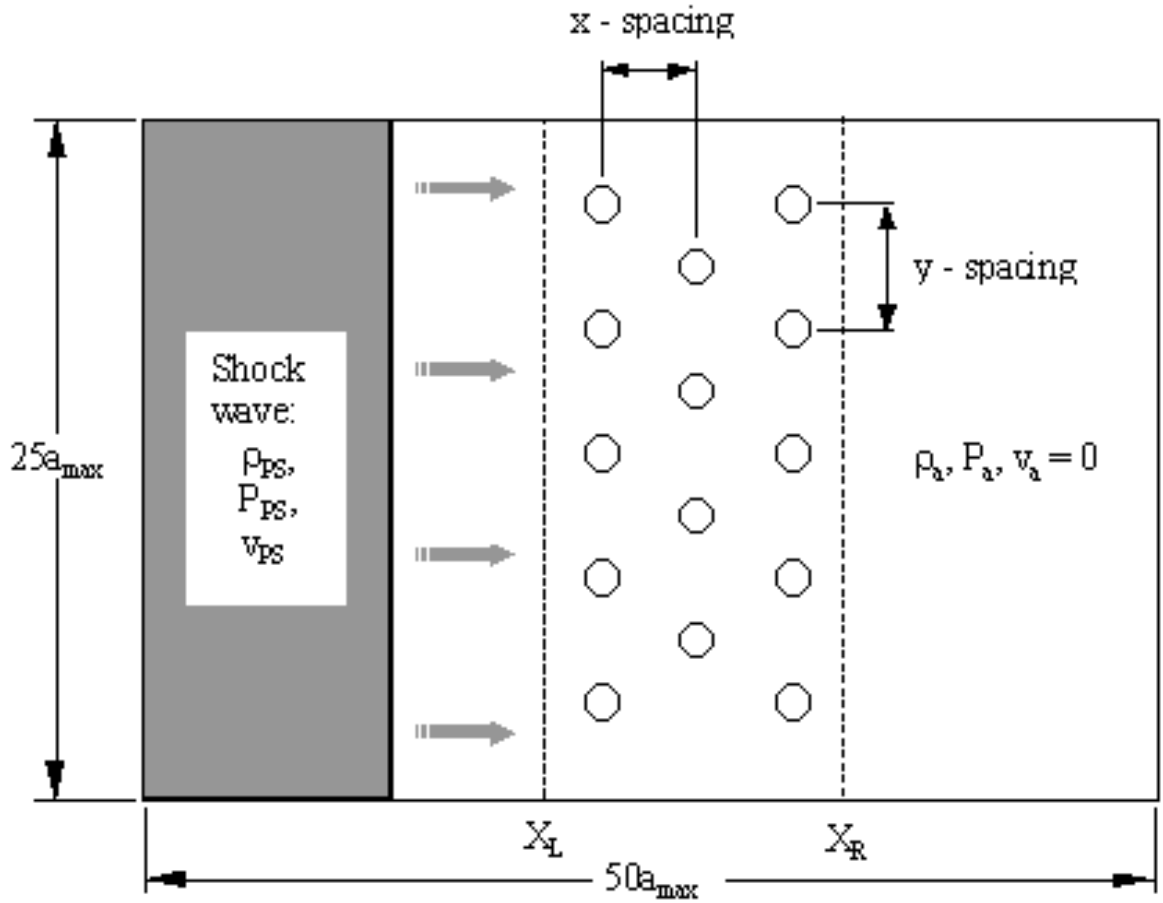}
\caption{Setup of the computational domain. Shown is the setup for the run
$M14$. ``x-spacing'' and ``y-spacing'' are the parameters used in Table~\ref{Runs}
for the description of the runs. Note: not drawn to scale.
\label{domain}}
\end{figure}

All our calculations were performed in a fixed reference frame in
which both the clouds and the ambient medium are stationary at
time $t=0$. In this reference frame the horizontal axis is taken
to be the x axis, the vertical axis - y axis. Initially both the
clouds and the surrounding intercloud medium are assumed to be in
pressure equilibrium and have pressure $P_{0}$. Typically, the
extent of the region, occupied by the cloud distribution at time
$t=0$, is taken to be not more than 30-35$\%$ of the horizontal
extent of the computational domain with $X_{L}$ offset by 5$\%$
from the left boundary of the computational domain and $X_{R}$
offset by 35-40$\%$. Table~\ref{Runs} below, describing the
numerical experiments discussed in this paper, provides the
details of the cloud distribution in each simulation.
Figure~\ref{domain} illustrates the setup of the computational
domain at $t=0$.

In the most general case we assume each cloud to have the same
nonuniform density profile. The clouds have constant density up to a
smoothing transition region at the cloud edge which is achieved
through a linear or $\tanh(r)$ function. We typically set the extent
of the transition region to the outer $20\%$ of a cloud radius $a_{i}$
and use the $\tanh(r)$ - type smoothing function. Therefore, the cloud
density profile is of the form
\beq
\rho_{i}(r)= \left\{ 
\begin{array}{ll}
\rho_{i}=Const, & 0\leq r \leq r_{i} \\
\displaystyle\frac{\rho_{a}+\rho_{i}}{2}+\frac{\rho_{a}-\rho_{i}}{2} \ \cdot
\ \frac{\tanh\big(r-\frac{a_{i}+r_{i}}{2}\big)}{\tanh\big(
\frac{a_{i}-r_{i}}{2}\big)}, & \ r_{i}\leq r \leq a_{i}
\end{array} \right.
\eeq

Although there is very little observational data available
concerning the internal structure of embedded clouds this
particular choice of the density profile seems to be a
sufficiently good approximation to the real physical clouds and
inhomogeneities. Burkert and O'Dell \citep{ODell98} discussed the
evidence for the exponential density profile in the cometary knots
of NGC 7293 (Helix nebula) which is similar to the density profile
used by us.

In the simple adiabatic interaction of a cloud with a shock wave
there are two dimensionless parameters that completely define the
problem: Mach number of the blast wave, $M_{S}$, and the density
contrast between the cloud and the intercloud medium
\begin{equation}
\chi_{i}\equiv\frac{\rho_{i}}{\rho_{a}}.
\end{equation}

The range of values spanned by the density contrast $\chi$ can be
quite large and is the most important parameter of the problem.
For the astrophysical situations of interest this range can often
cover up to 5 orders of magnitude (from 10 to $10^{6}$),
presenting a significant challenge both for the numerical modeling
and for the subsequent interpretation and analysis of the results.
In order to decrease the extent of this dimension of the parameter
space, we chose a ``compromise'' value of the parameter $\chi$ to
be 500. Although the runs we discuss in this paper all use this
value of the density contrast, we will briefly discuss numerical
experiments with $10 \lesssim \chi \lesssim 1000$ in the results
section, particularly in the context of the problem of mass
loading. We will provide a more comprehensive study of scaling
with density contrast in the subsequent work.

Another important parameter is the shock wave Mach number $M_{S}$.
We consider a planar steady shock wave propagating into the
computational domain from the left. Since we operate in the
reference frame in which both the clouds and the ambient medium
are stationary, the shock wave Mach number completely defines the
shock velocity as well as the conditions of the postshock flow.
Using the sound speed of the ambient medium
$C_{a}=(\gamma{P_{0}}/\rho_{a})^{1/2}$, the shock velocity in the
stationary reference frame takes the form
\begin{equation}
v_{S}=M_{S}\Big(\frac{\gamma{P_{0}}}{\rho_{a}}\Big)^{\frac{1}{2}}.
\label{vs}
\end{equation}
Using the Rankine-Hugoniot relations \citep{Landafshitz} we have the
following expressions for the postshock conditions\footnote{In our
discussion we assume the perfect gas, i.e. $\gamma=Const=\frac{5}{3}$
for cloud, intercloud, and postshock material.}
\begin{equation}
\rho_{PS}=\rho_{a}\bigg(1-\frac{2}{\gamma+1} \Big(1-\frac{1}{M_{S}^{2}}
\Big)\bigg)^{-1},
\label{rhops}
\end{equation}
\begin{equation}
P_{PS} = \displaystyle P_{0}\bigg(1+\frac{2\gamma}{\gamma +1}
\Big(M_{S}^{2}-1\Big)\bigg)=\displaystyle \frac{2}{\gamma +1}\rho_{a}
v_{S}^{2}\bigg(1-\frac{\gamma -1}{2\gamma M_{S}^{2}}\bigg),
\label{pps}
\end{equation}
\begin{equation}
v_{PS}=\frac{2v_{S}}{\gamma+1}\Big(1-\frac{1}{M_{S}^{2}}\Big).
\label{vps}
\end{equation}
We assume that the shock wave is strong, so that the condition
\begin{equation}
\Pi=\bigg(\frac{P_{PS}-P_{0}}{\rho_{a} C_{a}^{2}}\bigg)\gg 1
\label{pi}
\end{equation}
is satisfied.

Shock wave Mach numbers in astrophysical situations can cover a large
range of values. Fortunately, the problem becomes practically
independent on the Mach number for strong shocks, i.e.  for
$M_{S}\approx 10$ and above. Indeed, as it can be seen from the shock
conditions (\ref{rhops}) - (\ref{vps}), for $M_{S}\lesssim 10$ the
postshock density $\rho_{PS}$, pressure $P_{PS}$, and velocity
$v_{PS}$ are at most within a few percent of their respective values
at $M_{S}\rightarrow \infty$. Moreover, recalling that there is scale
invariance inherent in the hydrodynamic equations under
transformations
\begin{equation}
t\rightarrow tM_{S}, \quad v\rightarrow v/M_{S}, \quad P\rightarrow
P/M_{S}^{2},
\label{machscaling}
\end{equation}
the conclusion follows that for $M_{S}\gg 1$ the time evolution of a
cloud does not depend on the Mach number of the shock
\citep{KMC}\footnote{This conclusion is true with a restriction that the
shock speed $v_{S}$ is held fixed.}. Indeed, the results of KMC show
that for the difference in $M_{S}$ of 2 orders of magnitude ($10 -
1000$) time evolution of the system does not differ by more than
15$\%$. We will see that our results fully corroborate the presence of
Mach scaling in the problem under consideration.

We assume that the structure of the postshock flow does not change in
time for the duration of the simulations. An example of such steady
postshock flow is the wind from a post-AGB star driving a shock with a
constant postshock flow structure into a slow wind ejected during the
previous stages of evolution. This frees us from having to use the
pressure variation timescale $t_{P}$ to constrain a cloud size, since
we can set $t_{P} \rightarrow \infty$. On the other hand, for blast
waves one cannot assume a steady time independent postshock flow (for
example, SNR blast waves) and the size of the clouds is constrained by
the condition $t_{CC} \ll t_{P}$ as discussed by KMC (see \citep{KMC}).

It should be mentioned that the maximum cloud size is still
constrained by the condition of the shock front planarity. This
condition is less restrictive than the one discussed above,
however it still requires a cloud diameter not to exceed $5-10\%$
of the global shock wave front radius. This condition is
satisfied, for example, in the case of the inhomogeneities, or the
cometary knots, observed in such planetary nebulae as NGC 2392 and
NGC 7293 (e.g. \citep{ODell98}).

The timescale we use to define time intervals in our numerical
experiments is the time required for the incident shock wave to
sweep across an individual cloud, called the \emph{shock-crossing
time,}
\begin{equation}
t_{SC}=\frac{2a_{max}}{v_{S}},
\label{tsc}
\end{equation}
where $a_{max}=a_{0}$ for cloud distributions with identical clouds
and $a_{max}=max(a_{i})$ for cloud distributions of varying size
clouds.

Due to the scale-invariance of our simulations, one can, using
specific values for the shock velocity and the size of the
inhomogeneities, easily convert the time units used in our discussion
into the physical ones. $t_{SC}$ is particularly useful to
characterize the problem since it has clear physical meaning and does
not depend on a specific density contrast, which is important in the
case of systems containing clouds of different density.

Note that except for the very short period of time when a cloud
interacts with the shock front, the former finds itself immersed
in a post-shock flow or ``wind'' the pressure and density of which
vary only by several percent over the large range of Mach numbers.
Since KMC showed that the initial interaction with the shock front
does not alter the evolution of the system for the varying Mach
number, the details of the evolution should not change after the
shock front passed the cloud. Therefore, conclusions about Mach
scaling should be valid both for the durations of cloud-wind
interactions discussed by KMC, and for the much longer durations
in our experiments.

One final remark should be made concerning the boundary conditions
used in our experiments. In all runs we imposed a constant inflow
at the left boundary, described by the postshock conditions, which
is determined using the relations (4)-(6), and open boundary
conditions at the right, top, and bottom boundaries. Those outflow
boundary conditions were implemented via 0-order extrapolation.

\subsection{Description of the Runs}

All of the runs discussed in this paper contain a Mach 10 shock
wave as a part of the initial conditions and embedded clouds with
the density contrast of 500. Table~\ref{Runs} presents a summary
of our numerical experiments.

In addition to the dependence on the shock Mach number and the cloud
density contrast there are other degrees of freedom present even in
the simplest adiabatic case. We considered how the dynamical
evolution, e.g. rate of momentum transfer from the shock wave and
shock deceleration, mass loading, mixing of cloud material, etc. of
the system depend on

$\bullet$ the number of clouds present in the system;

$\bullet$ total cloud mass;

$\bullet$ spatial arrangement of clouds;

$\bullet$ individual cloud sizes and masses.

In most of the runs we constrained ourselves to the case of identical
clouds, varying only their number and arrangement. Radii of the clouds
in all runs except $M14_{r}$ is 2$\%$ of the horizontal extent of the
computational domain. In order to simplify consideration of the
dependence on a specific cloud arrangement, most runs have a regular
cloud distribution, where the clouds are placed in the vertices of the
mesh, formed by the centers of the clouds in the run M14. In addition,
we considered a more general case of a random cloud distribution with
random cloud spatial positions and radii.

All of our numerical experiments were run for about 100 $t_{SC}$
\footnote{For comparison, the experiments considered in KMC, that have
comparable initial cloud - ambient medium density contrast, were run
for about 25 $t_{SC}$.}. By this time each individual cloud has almost
completely lost its identity and gained a velocity comparable to the
velocity of the global flow. Mixing of cloud material with the global
ambient flow is nearly completed by 100 $t_{SC}$ as well.

In order to facilitate our analysis, we track temporal evolution of
the global averages and one-dimensional spatial distributions of
several quantities, namely

$\bullet$ kinetic energy fraction, $\eta_{kin}=E_{kin}/E_{tot},$

$\bullet$ thermal energy fraction, $\eta_{term}=E_{term}/E_{tot},$

$\bullet$ volume filling factor $\nu.$

In order to obtain those quantities from the complex data structure of
the adaptive mesh simulations, we project the values of the state
vector from each grid of the AMR grid hierarchy onto a uniform grid
with the resolution of the highest refinement level and that is
associated with the computational domain. Such projection does not
cause loss of data or its precision. When this projection is done, we
define the global averages of the first two quantities above as
\begin{eqnarray}
\langle\xi\rangle_{2D} = \frac{\displaystyle\sum_{i=1}^{N_{i}}
\sum_{j=1}^{N_{j}}
\xi_{ij}}{N_{i}N_{j}}\approx \frac{\displaystyle\int_{x_{min}}^{x_{max}}
\int_{y_{min}}^{y_{max}}\xi(x,y)dxdy}{(x_{max}-x_{min})(y_{max}-y_{min})},
\label{avxi2d}
\end{eqnarray}
where $\xi$ stands for the quantity being considered, and $N_{i}$,
$N_{j}$ are the numbers of cells of the projected grid in the $x$
and $y$ direction correspondingly. Such averaging allows us to
follow momentum transfer from the shock wave to the system of
clouds, in the case of $\langle\eta_{kin}\rangle_{2D}$, and
heating of the cloud system and intercloud material, in the case
of $\langle\eta_{term}\rangle_{2D}$.

We also define the one-dimensional spatial averages of those two
quantities as
\begin{equation}
\langle\xi\rangle_{1D}(x)=\frac{\displaystyle\sum_{j=1}^{N_{j}}
\xi_{ij}}{N_{j}}\approx\frac{\displaystyle\int_{y_{min}}^{y_{max}}
\xi(x,y)dy}{(y_{max}-y_{min})},
\label{avxi1d}
\end{equation}
where $\xi$ again stands for the quantity under consideration.

Our code follows advection of a passive tracer marking cloud
material. In order to follow mixing of the cloud material with the
global flow, we define the global average of the volume filling
factor $\langle\nu\rangle_{2D}$ as the ratio of the total number
of cells containing cloud material to the total number of cells in
the computational domain. We also define the one-dimensional
spatially averaged volume filling factor $\langle\nu\rangle_{1D}$
as the variation with the coordinate $x$ of the ratio of the
number of cells containing cloud material in each vertical row of
the computational grid to the total number of cells in the
vertical dimension.

\section{RESULTS}

\subsection{General Properties of the Shock-Cloud Interaction}

\begin{figure}
\epsscale{0.65}
\plotone{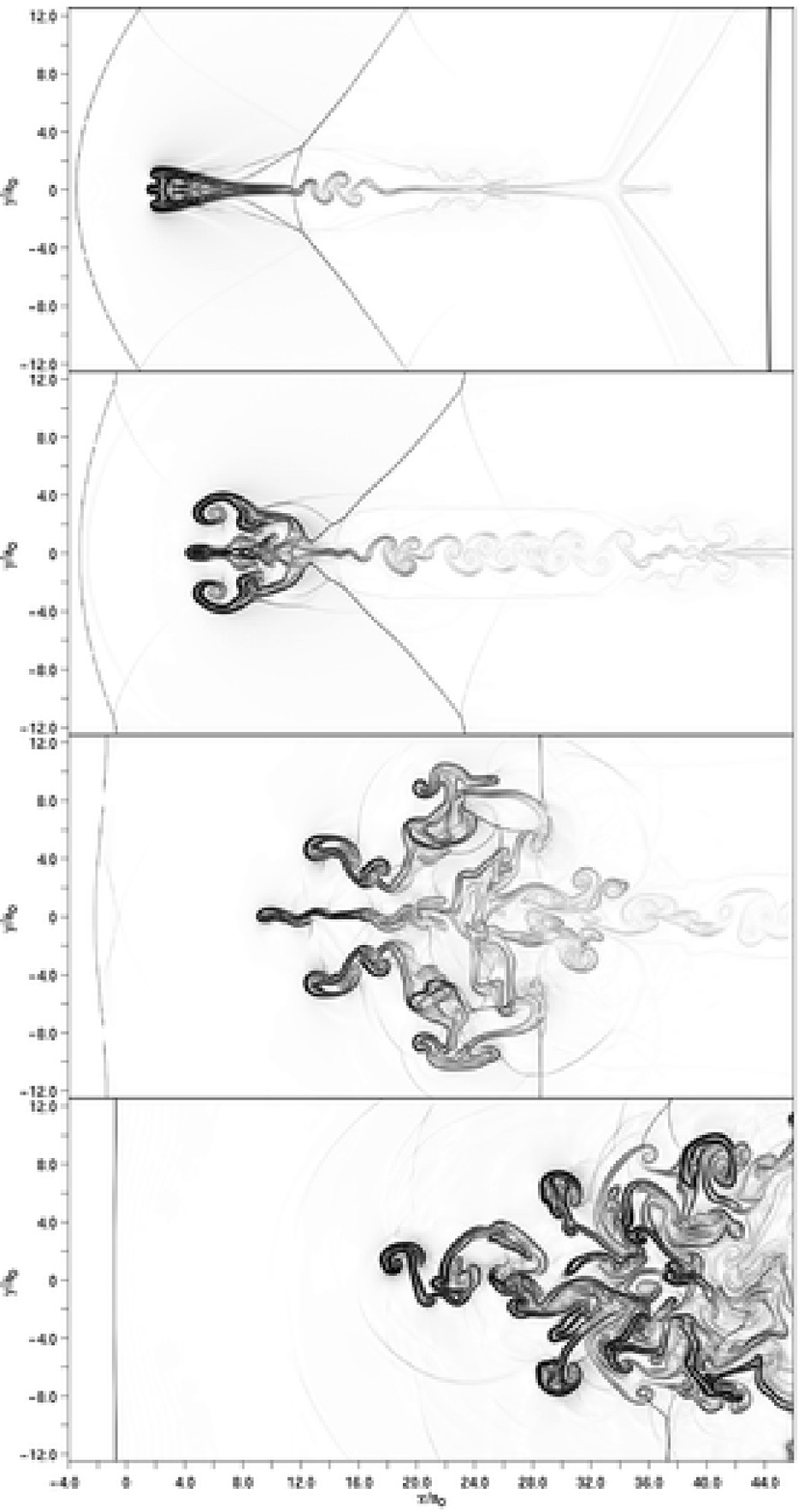}
\caption{Run $M1$. Time evolution of a system, containing a
single cloud and interacting with a $M_{S}=10$ shock wave. Shown are
the synthetic Schlieren images of the system at times $22.47 \
t_{SC}$, $35.23 \ t_{SC}$, $50.54 \ t_{SC}$, $68.40 \ t_{SC}$.
\label{1clump}}
\end{figure}

\begin{figure}
\epsscale{0.65}
\plotone{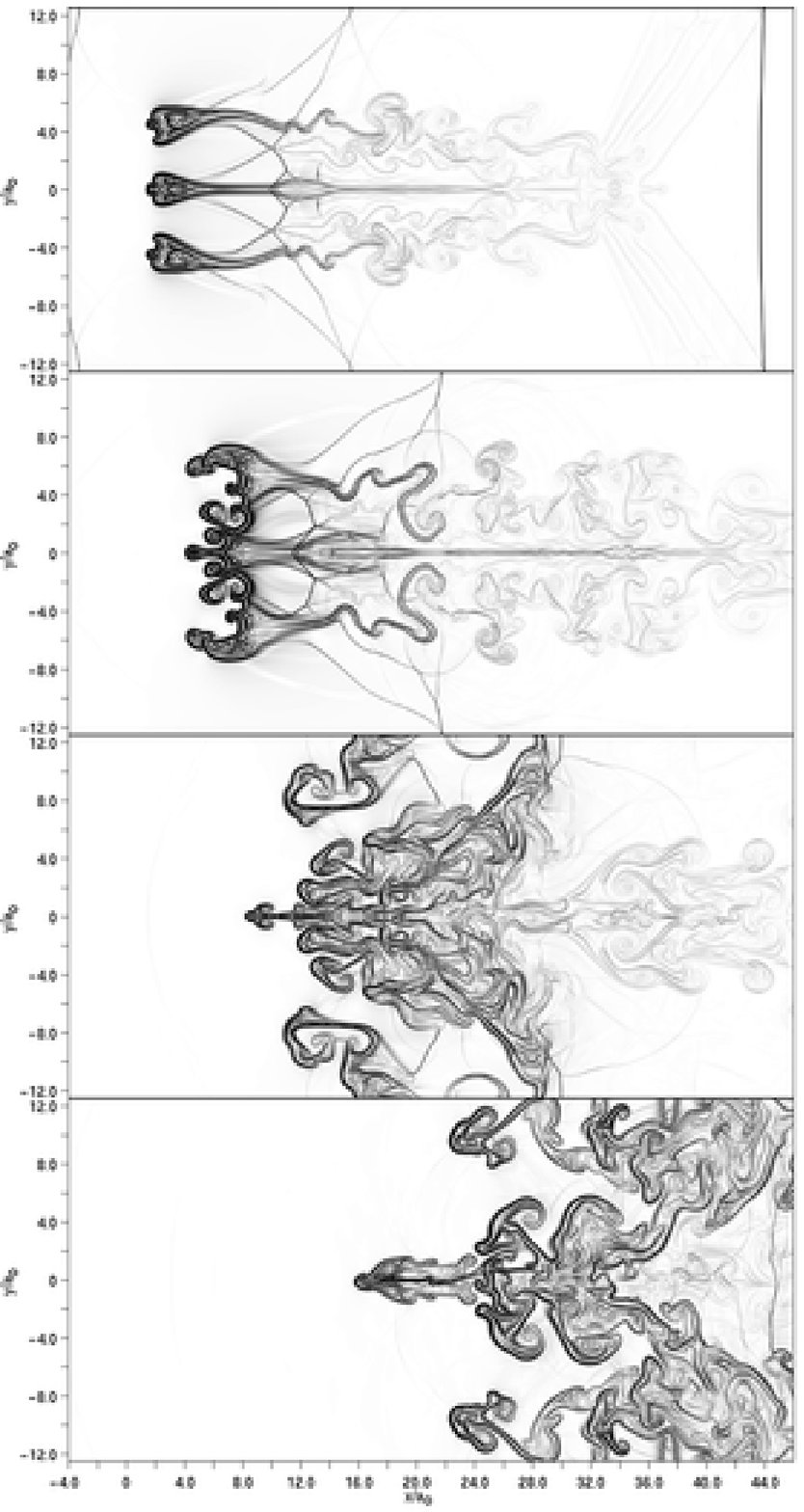}
\caption{Run $M3$. Time evolution of a system, containing three
identical clouds and interacting with a $M_{S}=10$ shock wave. Shown
are the synthetic Schlieren images of the system at times $22.47 \
t_{SC}$, $35.23 \ t_{SC}$, $50.54 \ t_{SC}$, $68.40 \ t_{SC}$.
\label{3clump}}
\end{figure}

Figures~\ref{1clump} - \ref{14random} show the time evolution of a
shock wave interacting with a single cloud (run $M1$), three
clouds (run $M3$), fourteen identical clouds in the regular
distribution (run $M14$), and fourteen clouds of random size in a
random distribution (run $M14_{r}$). Shown are the synthetic
Schlieren images of the system at four different times for all
four sequences. Each image is obtained by calculating the density
gradient at each point\footnote{To be more precise, the calculated
quantity is the gradient of the density logarithm. This makes the
images clearer and easier to understand.}, plotted on a gray scale
with the white denoting zero and black - the maximum density
gradient. Every image in each sequence roughly illustrates
transitions between the evolutionary phases discussed below.

\subsubsection{Initial Compression Phase}

After initial contact, an external shock transmits an internal
forward shock into a cloud. This causes cloud compression
and heating. At the same time a bow shock forms around the cloud.
KMC subdivide this phase into two stages: initial transient and
shock compression. Our numerical experiments show that, in
general, their description is applicable for all cloud
distributions except for the cases when individual clouds are
almost in contact at time $t=0$. However, we do not typically see
a reverse shock propagating inside the cloud as they did at later
stages. The cloud interior seems to be dominated by the forward
shock wave which prevents a reverse shock from detaching from the
back surface of the cloud. The absence of the reverse shock is the
reason for lower maximum densities in the cloud interior during
this compression phase compared with KMC: we typically see
$\rho_{i,max} \lesssim 3\rho_{i,0}$ as opposed to $\rho_{i,max}
\lesssim 10\rho_{i,0}$, quoted by KMC. Figure~\ref{shocks}
illustrates the major flow structures present in the system during
the initial compression phase.

Propagation of the forward shock in the cloud allows us to define
another important time scale governing the evolution of the system
and defining the duration of the compression phase: \emph{the
cloud crushing time} $t_{CC}$ \footnote{This was the principal
time scale in the study of KMC, although they defined it as as the
time necessary for the internal forward shock to cross the cloud
\emph{radius}. We have changed the definition in our work since
the definition of KMC did not actually correspond to the duration
of the compression phase. Therefore, $t_{CC}$ in our work is about
twice the $t_{CC}$ defined by KMC.}. This is the time necessary
for the internal forward shock to cross the cloud and reach its
downstream surface \beq t_{CC}=\frac{2a_{max}}{v_{CS}}. \eeq In
the above expression $v_{CS}$ is the internal forward shock
velocity and $a_{max}$ is again defined as $a_{0}$ in the cases of
cloud distributions with identical clouds, and as $max(a_{i})$ in
the cases of cloud distributions with clouds of varying size.
Following KMC, the velocity of the internal forward shock can be
written as \beq v_{CS} \simeq \frac{v_{S}}{\chi^{1/2}}\Bigg(
\frac{\gamma + 1}{8/3}
 \Bigg)^{1/2} \big( F_{c1}F_{st}\big)^{1/2},
\label{vcs} \eeq where $v_{S}$ is the velocity of the external
shock. The factor $F_{st}$ relates the external postshock pressure
far upstream with the stagnation pressure at the cloud stagnation
point and has the form \citep{KMC} \beq F_{st} \simeq
1+\frac{2.16}{1+10.7\{(\gamma + 1)\chi \}^{-1/2}}. \label{fst}
\eeq The factor $F_{c1}$ relates the stagnation pressure with the
pressure just behind the internal forward shock and has an
approximate value of 1.3 determined from numerical experiments
\citep{KMC}.

While we will primarily use the shock-crossing time as the major
time scale, we will occasionally give time in terms of the cloud
crushing time to facilitate comparison with the results discussed
by KMC. For this purpose we express the cloud crushing time in
terms of the shock crossing time. Recalling the definition of
$t_{SC}$ (9) we have \beq t_{CC}=\Bigg( \chi^{1/2}\Big(
\frac{8/3}{\gamma + 1} \Big)^{1/2} \big(
F_{c1}F_{st}\big)^{-1/2}\Bigg)t_{SC}. \eeq Therefore, for the case
of $\chi=500$ \beq t_{CC} = 12 \ t_{SC}, \label{tcc} \eeq which
agrees to about a few percent with the results of the numerical
experiments.

The global properties of the flow at this stage are characterized
by the onset of individual bow shocks around each cloud in a time
of order $t_{SC}$. By the end of the initial compression phase
those individual bow shocks merge into a single bow shock.
\footnote{It should be noted that a \emph{bow wave} forms instead
of a bow shock if the external postshock flow is subsonic, i.e. if
\begin{displaymath}
M_{PS}=v_{PS}\bigg(\frac{\gamma
P_{PS}}{\rho_{PS}}\bigg)^{-\frac{1}{2}} \leq 1.
\end{displaymath}
With the postshock conditions $\rho_{PS}$, $P_{PS}$, and $v_{PS}$
determined from the relations (\ref{rhops}) - (\ref{vps}), the above
criterion is satisfied for the following values of the external shock
Mach number
\begin{displaymath}
M_{S} \leq \bigg( \frac{-\beta+\sqrt{\beta ^{2}-4\alpha
\delta}}{2\alpha}
\bigg)^{\frac{1}{2}} \approx 2.758 \ for \ \gamma=\frac{5}{3},
\end{displaymath}
where
\begin{displaymath}
\alpha=4-2\gamma(\gamma -1), \ \beta = \gamma^{2}-6\gamma-7, \
\delta = 2\gamma + 2.
\end{displaymath}
Since in this paper we consider the external shocks, Mach numbers of
which are typically above 5.0, we will hereafter not consider the
possibility of a bow wave formation.}

Finally, the downstream flow, i.e. the flow right behind the
external forward shock front, is effected by the onset of
turbulence in the tails behind the clouds.

\subsubsection{Re-expansion Phase}

This phase is initiated after the cloud internal forward shock
reaches the back of the cloud. The two major processes then occur:
lateral expansion of the cloud and the onset of instabilities at
its upstream surface. At this stage Rayleigh-Taylor type
instabilities dominate at the cloud/ambient flow interface.  These
are driven in part by the cloud expansion and incipient
large-scale fragmentation. The flow downstream with respect to the
clouds is dominated by Kelvin-Helmholtz instabilities operating in
the growing turbulent region. The combined action of the lateral
expansion and the instabilities causes the clouds to take the
``umbrella-type'' shape and eventually break up.

In the context of those two processes, the initial cloud
separation becomes of key importance defining the subsequent
behaviour of the whole system. We will see below that it can be
used to distinguish between the two regimes of cloud evolution:
interacting and noninteracting, and can serve as the basis for
classification of cloud distributions. In subsection 3.3 we will
give more rigorous discussion of the role of cloud separation. For
now we give a qualitative illustration.

Clouds, located far enough from each other, are not greatly
influenced by their neighbours and their interaction with the flow
proceeds independently as described by KMC. This case is
illustrated in Figure~\ref{2clumpwide}. Compared to the evolution
of a single cloud system, shown in Figure~\ref{1clump}, the two
clouds evolve up to the point of their destruction very similarly
to the single cloud case. However, cloud separations can be small
enough for the mutual interaction to manifest early during the
re-expansion phase, as in Figure~\ref{2clump}. This mutual
interaction causes changes primarily in the flow between clouds.
As a result the lateral expansion and growth of the
Rayleigh-Taylor instabilities in the cloud material is affected.
The tails behind the clouds are also deformed outwards (see, for
example, as well Figure~\ref{3clump}).

The unperturbed supersonic flow that forms behind the external shock
wave undergoes a transition from a supersonic to a subsonic regime as
it passes through a cloud bow shock.  As a consequence it suffers a
significant velocity drop whose magnitude is larger for smaller cloud
separations due to the larger volume of the stagnation zone in front
of the clouds. Clouds, acting as the Lavalle nozzles, then cause the
flow material to re-accelerate.  The flow reaches a sonic point next
to a cloud core for the regions of the flow adjacent to a cloud, and
further downstream for the regions of the flow located further from
the clouds.  It is important to note that this re-acceleration results
in rarefaction of the flow and a gradual decrease both of
thermodynamic and dynamical pressure. Eventually, as a result of
acceleration in the intercloud region, the flow becomes highly
supersonic and finally shocks down through a stationary shock formed
downstream of the clouds to the regime close to the unperturbed flow
behind the external shock (see Figures~\ref{2clumpwide}-\ref{2clump}).

From the above discussion it is clear that the lateral expansion
velocity depends critically on the cloud separation. For
sufficiently low flow speeds the cloud material will expand at the
cloud internal sound speed. With increasing global flow velocities
(or, equivalently, with increasing velocities of the external
shock front) the lateral expansion velocity will increase as well.
This velocity is limited, in principle, by the terminal expansion
velocity into vacuum.

For a fixed unperturbed upstream flow, the flow velocity near a cloud
lateral surface (facing the space in between the clouds) will be the
highest in the case of a single cloud or a cloud located far from the
neighbouring ones. With decreasing cloud separation this velocity will
decrease as well, causing higher dynamical pressure on the lateral
surface and, therefore, lower lateral expansion velocities.  This
occurs because the velocity drop across a bow shock in the cases of
small cloud separations is much larger due to a stronger stagnation
effect in between the bow shock and the clouds. Therefore, flow
adjacent to the cloud does not reach velocities as high as in cases of
large cloud separations \footnote{It should be noted that eventually
the velocities reached by the flow downstream after passing the region
between clouds are much higher and, consequently, the strength of the
stationary shock downstream is much larger in the case of small cloud
separations.}. Another way to look at this process is the
following. The flow adjacent to the cloud surface passes through a
sonic point but, in the cases of small cloud separations, densities in
the stagnation region are much higher.  Thus flow densities at the
sonic point near the cloud lateral surface are much higher. This leads
to lower sound speeds and, therefore, lower flow speeds.

Following KMC, the \emph{effective lateral expansion velocity}
$v_{exp}$ can be defined as the internal cloud sound speed 
\beq
v_{exp}=C_{C}=v_{CS}\frac{\big(2\gamma(\gamma -
1)\big)^{\frac{1}{2}}}{\gamma + 1}, 
\label{cc} 
\eeq 
where $v_{CS}$ is the velocity of the cloud internal forward shock
(\ref{vcs}). Our numerical experiments prove this to be a very good
approximation during almost all of the re-expansion phase. The
expansion velocity exceeds this value by the end of the re-expansion
phase due to stagnation pressure in the regions, formed by the
Rayleigh-Taylor instability.

We are now in a position to articulate the temporal evolution of a
cloud radius in the direction perpendicular to the upstream flow.
From the moment of their initial contact with the external shock to
the moment of their destruction, the clouds first undergo slight
compression in the direction perpendicular to the flow and
subsequently re-expand. KMC's analytic model did not explicitly
include cloud compression but instead tried to account for its effect
via a reduced monotonic expansion rate from $t=0$. Since
$a_{\perp}(t)$ is intimately related to the drag exerted on a cloud by
the global flow, the theoretical rate of the momentum pickup by a
cloud (or the rate of cloud deceleration in the reference frame used
by KMC) differed from the numerical result.  Namely, in Figure 12b of
the paper by KMC numerical and theoretical results are practically the
same up to the time $\approx 2.0t_{CC}$, when the rate of cloud
deceleration suddenly increases and the numerical and theoretical
results drastically diverge. This moment of time corresponds to the
beginning of the re-expansion phase, when the cloud cross-section
starts to increase causing an increase of the rate of the momentum
transfer from the flow to the cloud. To avoid this problem and
simplify an expression for $a_{\perp}(t)$ we use the following form
for evolution of a cloud radius normal to the flow,
\beq 
a_{\perp}(t)= \left\{ \begin{array}{ll} a_{0}, & t \leq t_{CC} \\
a_{0}+C_{C}(t-t_{CC}), & t_{CC} \leq t \leq t_{CD}.
\end{array} \right.
\label{aperp}
\eeq
Here, $C_{C}$ is given by (\ref{cc}), and $t_{CD}$ is the cloud
destruction time, defined below in (\ref{tcd}).

\subsubsection{Cloud Destruction Phase}

Depending on the cloud separation, via the process of re-expansion
clouds may come into contact and merge into a single coherent
structure.  This subsequently interacts with the flow as a whole and
eventually breaks up. Thus for the case of small cloud separations we
can define the moment of cloud merging as the onset of the cloud
destruction phase. For large cloud separations in which individual
clouds get destroyed before ever merging, it is difficult to define
the precise onset of the destruction phase as it may be effectively
viewed as a part of the re-expansion phase.

We define the end of the cloud destruction phase as the time when the
largest cloud fragment contains less than 50$\%$ of the initial cloud
mass. For single cloud systems or systems of weakly interacting clouds
we define the total time from $t=0$ until the end of the cloud
destruction phase as the \emph{cloud destruction time} $t_{CD}$,
\beq t_{CD}=\alpha t_{CC} = \alpha' t_{SC}. 
\label{tcd} 
\eeq
Typically $\alpha \approx 2.0$ consistent with KMC, and using
(\ref{tcc}) we find $\alpha' \approx 24$ in our simulations.

In addition to $t_{CD}$ there is also a \emph{cloud system destruction
time} $t_{SD}$ which we define as the time when the largest fragment
of a cloud located furthest downstream contains less than 50$\%$ of
its initial mass. For thick layer systems (to be described later),
including strongly interacting cloud distributions, $t_{CD}$ becomes
less relevant as a description of the system than $t_{SD}$ because
$t_{CD} \ < \ t_{SD}$.

\subsubsection{Mixing Phase}

After the end of the destruction phase cloud material velocity is
still only a small fraction of the global flow velocity (see
(\ref{vcmax}) below). The velocity difference promotes
Kelvin-Helmholtz instabilities at the cloud material - global flow
interfaces and, therefore, the transition of the system to a turbulent
regime. Typically by the beginning of this phase each cloud has lost
its identity as a result of merging with neighbouring clouds. As the
individual fragments become smaller and the velocity of the global
flow relative to the cloud material decreases, Kelvin-Helmholtz
instabilities grow faster than the Rayleigh-Taylor type ones. This
eventually results in complete domination by the former of the
small-scale fragmentation and causes mixing of cloud material with the
flow \citep{KMC}.

In our numerical experiments, as it can be seen in
Figures~\ref{1clump} - \ref{14random}, turbulent mixing produces a
two-phase filamentary system. The appearance of such a two-phase
system results because our code does not include viscous diffusion
or thermal conduction restricting all dissipative effects to
numerical diffusion only. The latter acts at the length scales
comparable with a cell size at the highest refinement level.

Real dissipative effects also constrain the overall stability of
cold dense plasma embedded in a tenuous hotter medium. KMC
considered the overall effect of thermal conduction on the
stability of such two-phase media against evaporation. They
concluded that the cloud ablation time due to evaporation,
expressed in terms of the shock-crossing time $t_{SC}$, defined in
(\ref{tsc}) above, has the form 
\beq
t_{ab}=\frac{\chi}{9F(\sigma'_{0})}\Bigg(\frac{2}{\gamma+1}
\Bigg)^{\frac{1}{2}}\Big(F_{c1}F_{st}\Big)^{-\frac{1}{2}}t_{SC},
\label{tablation} 
\eeq 
where $F(\sigma'_{0})$ is typically of order unity
\citep{KMC}. Therefore, for the case of our simulations, the typical
ablation time is $t_{ab} \sim 30t_{SC}$, or comparable to the cloud
destruction time.

One can also estimate an effective depth over which diffusion and
thermal conduction will disrupt the boundary layer between the two
phases over a dynamical time-scale, $t_{SC}$. This can be
estimated as follows (see \citep{Blackman96} and references
therein). For viscous diffusion 
\beq
d_{diff} \sim (D_{diff}t_{SC})^{\frac{1}{2}} = \Bigg(\bigg(\frac{2}
{\gamma}\bigg)^{\frac{1}{2}}\frac{1}{M_{S}n_{0}\sigma}a_{0}\Bigg)^{
\frac{1}{2}},
\label{ddiff}
\eeq where $D_{diff}$ is the diffusion coefficient and $n_{0}$ is 
the initial cloud number density. If we assume $n_{0} \sim 1000 \
cm^{-3}$ then, for the cases presented in our simulations, $d_{diff}$
is about $1\%$ of the initial cloud radius, or equivalently is about
$1/5$ of a cell of the computational domain at the highest refinement
level. For thermal conduction, the effective depth is
\beq
d_{term} \sim (D_{term}t_{SC})^{\frac{1}{2}} = \Bigg(\bigg(\frac{2}
{\gamma}\bigg)^{\frac{1}{2}}\frac{1}{M_{S}n_{0}\sigma_{p}}\bigg(
\frac{m_{p}}{m_{e}}\bigg)^{\frac{1}{2}}a_{0}\Bigg)^{\frac{1}{2}}
=d_{diff}\bigg(\frac{m_{p}}{m_{e}}\bigg)^{\frac{1}{4}}.
\label{dterm}
\eeq
Then $d_{term}$ is about $6\%$ of $a_{0}$, or equivalently, about
twice the size of a computational cell at the highest refinement
level. Therefore, should we have included real dissipative effects
they would destroy the smallest resolvable structures over the
dynamically relevant time scales. Consequently, any further increase
in resolution without providing for the appropriate mechanisms,
capable to inhibit significantly diffusion and thermal conduction,
would not provide additional insights into the real physical evolution
of a system.

The importance of the dissipative effects is two-fold. First
consider the stability of the initial system against destruction
due to thermal conduction and diffusion. From the arguments given
above dissipative effects prevent survival of the system for any
dynamically significant amount of time. As a solution to this
problem, KMC suggested that weak magnetic fields inhibit thermal
conduction and diffusion. Indeed, as it was shown by
\citep{MacLow94}, evolution of weakly magnetized clouds during the
compression and re-expansion phases does not differ significantly
from the purely hydrodynamic description. However the presence of
magnetic fields would raise other issues. During the mixing phase
the system undergoes transition to turbulence which may amplify
the initially dynamically insignificant magnetic fields.
Turbulence can lower values of the plasma parameter $\beta =
P_g/P_B \rightarrow 1$ or even smaller, which may alter the
evolution of the system during the later periods of the mixing
phase. In this respect only a fully magnetohydrodynamic study of
the evolution of a system of clouds interacting with a strong
shock is fully self-consistent (for a series of single cloud MHD
studies see \citep{MacLow94}, \citep{Gregori99}, \citep{Gregori2000},
\citep{Jones96}, \citep{Miniati99}, \citep{Lim99}, \citep{Jun99}).

\subsection{Role of Cloud Distribution}

In order to characterize the global properties of the shock/cloud
system interaction we plotted the time evolution of the global
quantities defined in section 2.3 for the runs $M1$, $M2$, $M3$,
$A5$, $M14$, and $M14_{r}$. Those plots are presented in
Figures~\ref{ekin} - \ref{fill}.

The important feature of those plots is the striking similarity of
the behaviour of systems containing similar cloud distributions.
The systems containing from one to five clouds arranged in a
single layer exhibit exactly the same rate of momentum transfer
from the global flow. This is manifested by the linear rates of
fractional kinetic energy $\langle \eta_{kin} \rangle_{2D}$
increase from $t=0$ up to $t=24t_{SC}$ (see Figure~\ref{ekin}).
The value of the slope for those five cases is $0.193\pm 1.6\%$.
The thermal energy $\langle \eta_{term} \rangle_{2D}$ behaves
complementarily (see Figure~\ref{eterm}). Such behavior of single
layer systems contrasts that of the multiple layer systems, namely
the runs $M_{14}$ and $M_{14r}$, which we now discuss.

The two fourteen cloud runs have different cloud distributions
(regular as opposed to random), different total cloud mass and
different cloud sizes. Nevertheless, the evolution of their
fractional energies are similar. The rate of the kinetic energy
increase during compression and re-expansion is the same for both
$M_{14}$ and $M_{14r}$ and yet is different from that single layer
cases.  The slope in the multi-layer cases is also practically
constant throughout the two phases with values $0.146 \pm 4.5\%$.

Note that for all cases the kinetic (thermal) energy reaches its
maximum (minimum) at the time $t=24t_{SC}$, or the time, defined above
as the cloud destruction time $t_{CD}$, even though for the fourteen
cloud runs the cloud system destruction time $t_{SD}$, defined above
in subsection 3.1.3, is greater than $t_{CD}$. Thus it seems
reasonable to conclude that the cloud destruction time is a universal
parameter independent on the details of a cloud distribution.

After passing through its maximum, the kinetic energy fraction begins
to decrease due to the transition to turbulence and, consequently,
turbulent energy dissipation. It is difficult to define a value of the
slope for the mixing phase due to the complex nature of the turbulent
flow but the average rate of kinetic energy dissipation in the system
is $\approx 0.013 \pm 27\%$ for the single layer systems and $\approx
0.015 \pm 25\%$ for the multiple layer ones. The proximity of these
two values (within the standard error) is evidence that the systems
have lost any unique details of the initial cloud distribution and
developed turbulence that depends primarily on the rate of energy
input at the largest scale, i.e. on the relative velocity of the
global flow with respect to the cloud material.

The similarity in behaviour of single vs multiple layer systems is
even more prominent in the time evolution of volume filling
factors $\langle \nu \rangle_{2D}$. As can be seen in
Figure~\ref{fill}, the rate of cloud material mixing into the
global flow is the same for all single layer systems but is
different from that in multiple layer ones. The higher mixing rate
in the case of multiple layer distributions results because
upstream clouds pick up momentum faster than the downstream ones.
Upstream clouds promote destruction of the downstream ones and
consequently the overall mixing of the system.\footnote{The
presence of the maximum values in $\langle \nu \rangle_{2D}$ in
Figure~\ref{fill} for all runs is due to the eventual loss of the
cloud material through the outflow boundaries.}

These results lead us to conclude that cloud distribution plays a
more important role than the number of clouds or the total cloud
mass. We use this conclusion in the next section as the foundation
for classifying possible cloud distributions and defining the
general type of the cloud system evolution in each category.

\subsection{Critical Density Parameter}

We have seen that the cloud distribution plays the defining role
in determining the evolution of a shock-cloud system. We now
quantify this statement and define criteria for determining the
behaviour of a given system.

We define a set of all possible cloud distributions for a given
number of clouds $N$. We consider only the clouds of equal or
comparable size and density contrast. We define each set of cloud
distributions $\lambda_{N}$ for any given number of clouds $N$ to
be a set of all possible $N$ pairs of cloud center coordinates,
satisfying two conditions: (1) each pair of clouds is separated by
some minimum distance $r_{min}$ and (2) clouds are confined to a
layer extending from the position $X_{L}$ to the position $X_{R}$ (see
Figure~\ref{domain}):
\beq
\begin{array}{c}
\forall N\ge 1 : \lambda_{N}\equiv\{(x_{i},y_{i}),1\le i\le N\ :
r_{ij}={\big((x_{i}-x_{j})^2+(y_{i}-y_{j})^2\big)}^{\frac{1}{2}}\ge
r_{min}\approx 2a_{0}, \\ x_{i} \in \lbrack X_{L},X_{R}\rbrack \}.
\end{array}
\eeq

Next, we consider the complete set of all possible cloud distributions
$\Lambda$ for all possible cloud numbers defined as
\beq
\Lambda \ = \ \displaystyle\bigcup_{N\ge 1} \{\lambda_{N}\}.
\eeq
We define within this set $\Lambda$ the two subsets: a subset of
\emph{``thin-layer''} cloud distributions $\Lambda_{I}$ and a subset
of \emph{``thick-layer''} cloud distributions $\Lambda_{M}$ so that
\beq
\Lambda_{I} \ \cup \ \Lambda_{M} = \Lambda \quad and \quad \Lambda_{I}
\cap \ \Lambda_{M} \ \equiv \emptyset.
\eeq
In our numerical experiments those two subsets are associated with the
single row and multiple row distributions.

In order to give a precise definition of those two fundamental classes
of cloud distributions we need to introduce several auxiliary quantities.

\subsubsection{Cloud Velocity and Displacement}

We now estimate the distance that the cloud material will travel
before the cloud breakup, i.e. within the time $t_{CD}$.

The equation of motion of a cloud in the stationary reference
frame of the unshocked ambient medium takes the form
\beq
m_{C}\frac{dv_{C}}{dt}=\frac{1}{2}C_{D}\rho_{PS}\{v_{PS}-v_{C}\}
^{2}A_{C}(t),
\label{emotion}
\eeq
where $m_{C}$ is the mass of the cloud, $v_{C}$ is the cloud velocity
in the stationary reference frame, $C_{D}$ is the cloud drag
coefficient, $\rho_{PS}$ is the undisturbed postshock flow density,
and $A_{C}(t)$ is the cloud cross section area normal to the flow. It
should be noted that this equation is valid only until the cloud
destruction is complete, i.e. until $t\approx t_{CD}$. From this point
on we assume that the drag coefficient $C_{D}\approx 1$ which is a
rather good approximation for a cylindrical body embedded in a
supersonic flow of $M_{PS}=1.31$ (see KMC and \citep{Bedogni98}).

Let us assume for a moment that the clouds have finite extent in
the z-direction: $z_{0}$. Note that then
\beq
m_{C}=\rho_{0}\pi
a_{0}^2 z_{0},
\label{mass} 
\eeq
where $\rho_{0}$ is the cloud density and $a_{0}$, $z_{0}$ are the
cloud dimensions at time $t=0$. Moreover, the cross section area is
\beq
A_{C}(t)=2a_{\perp}(t)z_{0},
\label{crossa}
\eeq where
$a_{\perp}(t)$ is the cloud radius in the direction normal to the
flow. Substituting (\ref{mass}) and (\ref{crossa}) into
(\ref{emotion}) and using (\ref{aperp}) for $a_{\perp}(t)$ we get the
following modified equation of motion \footnote{Note, that from now on
we will omit the cloud drag coefficient considering it to be equal to
1.}
\beq
\frac{dv_{C}}{dt}=\frac{\rho_{PS}}{\rho_{0}\pi
a_{0}^2}\{v_{PS}-v_{C} \}^{2}a_{\perp}(t).
\label{newemotion}
\eeq
This equation describes motion of the cloud as a result of its
interaction with the postshock wind. However, we also need to account
for the velocity that the cloud material acquires after its initial
contact with the external shock front. This velocity may be comparable
to the velocity acquired during the compression and re-expansion
phases and, therefore, must be carefully included into consideration.

Recall that the initial contact of the incident shock front drives an
internal forward shock into the cloud with velocity $v_{CS}$.  Cloud
material behind the internal shock front gains a velocity $v_{C,PS}$,
that can be determined from the Rankine-Hugoniot relations in the
usual manner,
\beq
v_{C,PS}=\frac{2v_{CS}}{\gamma+1}\Big(1-\frac{1}{M_{CS}^{2}}\Big).
\label{vcps1}
\eeq
Here $M_{CS}$ is the Mach number of the cloud internal forward shock,
which can be expressed in terms of the external shock Mach number as
follows
\beq M_{CS}=\frac{v_{CS}}{C_{C,0}}=M_{S}\Bigg(\frac{\gamma +
1}{8/3} \Bigg)^{1/2} \big( F_{c1}F_{st}\big)^{1/2},
\label{mcs}
\eeq
where by $C_{C,0}$ we denoted the sound speed in the unshocked cloud
material and used (\ref{vcs}) for $v_{CS}$.  For the simulations
discussed in this paper ($\chi=500$ and $\gamma=5/3$) the internal
cloud shock Mach number is $M_{CS}=1.86M_{S}=18.6$.

Substituting (\ref{vcs}) for $v_{CS}$ and (\ref{mcs}) for $M_{CS}$
into (\ref{vcps1}), and expressing the external shock velocity
$v_{S}$ in terms of the unperturbed upstream postshock velocity
$v_{PS}$ by means of (\ref{vps}), we obtain the following
expression for the velocity of the cloud material due to the cloud
interaction with the external shock front
\beq
v_{C,PS}=v_{PS}\Bigg(\frac{\Big(\frac{\gamma + 1}{8/3}\Big)^{1/2}
\big( F_{c1}F_{st}\big)^{1/2}\Big(M_{S}^2-\frac{8/3}{\gamma + 1}
\big( F_{c1}F_{st}\big)^{-1}\Big)}{\chi^{1/2}\big(M_{S}^2-1\big)}\Bigg)
=v_{PS}\Gamma.
\label{vcps2}
\eeq
For the case of $M_{S}=10$, $\Gamma=0.084$. Note that for the limiting
case $M_{S} \rightarrow \infty$ the value of $\Gamma$ remains
practically unchanged at $0.083$ which corroborates the previously
discussed Mach scaling (see (\ref{machscaling})).

Finally, making use of the fact that the relative velocity of the
postshock flow with respect to the cloud is now
$(v_{PS}-v_{C,PS}-v_{C})$, we can integrate (\ref{newemotion}) and
obtain the following form of the cloud velocity
\beq
v_{C}(t)= \left\{ \begin{array}{ll}
\displaystyle v_{PS}\Big(1-\frac{1}{At + (1-\Gamma)^{-1}}\Big), &
 t \le t_{CC} \\
\displaystyle v_{PS}\Big(1-\frac{1}{AB(t-t_{CC})^2+A(t-t_{CC})+C}\Big),
& t_{CC} \le t \le t_{CD},
\end{array} \right.
\label{vc}
\eeq
where we introduced the following quantities
\beq
A = \frac{\rho_{PS} v_{PS}}{\rho_{0} \pi a_{0}} =
\frac{1}{t_{SC}\chi}\cdot \displaystyle\Bigg(\frac{\pi}{2}\Big(
\frac{2}{\gamma+1}\bigg(1-\frac{1}{M_{S}^2}\bigg)\Big)^{-1}-1\Bigg)^{-1};
\label{auxa} 
\eeq 
\beq B = \frac{C_{C}}{2a_{0}}=\frac{1}{t_{SC}\chi^{1/2}}\cdot \big(
F_{c1}F_{st}\big)^{1/2}\Bigg( \frac{3\gamma(\gamma - 1)} {4(\gamma
+ 1)}\Bigg)^{\frac{1}{2}};
\label{auxb}
\eeq
\beq C =12At_{SC}+(1-\Gamma)^{-1}.
\label{auxc}
\eeq
The unperturbed postshock quantities $\rho_{PS}$ and $v_{PS}$ are
determined from the conditions (\ref{rhops}) and (\ref{vps}), $C_{c}$
is the sound speed in the shocked cloud, defined by (\ref{cc}), and
the factor $F_{st}$ is defined by the relation (\ref{fst}).

The first quantity $A$ relates the specific momentum of the
postshock wind to the cloud inertia (mass). Thus it defines the
rate of the momentum pickup by a cloud during the compression
phase, when the cloud dimension transverse to the flow does not
increase. The second quantity $B$ is the inverse sound crossing
time in a compressed cloud, i.e. at the end of the compression
phase, again for the cloud dimension transverse to the flow. This
quantity determines the rate of the cloud lateral expansion.
Therefore, during the re-expansion phase the regular momentum
transfer from the wind to the cloud, described by A, is augmented
by the cloud lateral expansion, described by $B$, which comes as
an additional factor in the quadratic dependence on $t$. Quantity
$C$ ensures continuity of the cloud velocity during the transition
from the compression to the re-expansion phase.

Next, integrating (\ref{vc}) from time $t=0.0$ up to the cloud
destruction time $t=t_{CD}$ we can determine the displacement of cloud
material during the compression and re-expansion phases
\beq
L_{C}(t)= \left\{ \begin{array}{ll}
\displaystyle v_{PS}\Bigg(t-\frac{1}{A}ln\Big((1-\Gamma)At+1\Big)\Bigg),
& 0 \le t \le t_{CC} \\
v_{PS}\displaystyle\Bigg(t-\frac{1}{A}\bigg(\frac{2}{q}tan^{-1}
\frac{(t-t_{CC})q}{(t-t_{CC})+\frac{2C}{A}}+ln\Big((1-\Gamma)At_{CC}
+1\Big)\bigg)\Bigg), & t_{CC} \le t \le t_{CD}
\end{array} \right.,
\label{l}
\eeq
where $q=\sqrt{4BC/A-1}$. This allows us to estimate the total
displacement a cloud incurs before its destruction, called
\emph{the cloud destruction length},
\beq
L_{CD}=L_{C}(t_{CD}).
\label{lcd}
\eeq

In order to get a clearer understanding of the general expressions
(\ref{vc}) and (\ref{l}) let us consider the two cases: the case
presented in our simulations with $M_{S}=10$ and the limiting case of
$M_{S} \to \infty$. We will assume in both cases the density contrast
of $\chi=500.0$ and $\gamma = 5/3$.

First, we rewrite (\ref{vc}) as
\beq
v_{C}(t)= \left\{ \begin{array}{ll}
\displaystyle v_{PS}\Bigg(1-\Big(\frac{t}{t_{SC}}
a_{1}+ a_{2}\Big)^{-1}\Bigg), & t \le t_{CC} \\
\displaystyle v_{PS}\Bigg(1-\Big(\bigg(\frac{t}{t_{SC}}-12\bigg)^2b_{1}
+\frac{t}{t_{SC}}a_{1} + a_{2}\Big)^{-1}\Bigg),
& t_{CC} \le t \le t_{CD}
\end{array} \right.
\label{vcnum}
\eeq

In the first case of $M_{S}=10$ the coefficients $a_{1}$, $a_{2}$, and
$b_{1}$ have the following values
\beq
a_{1} \ = \ 1.79 \cdot 10^{-3}; \ a_{2} \ = \ 1.09; \
b_{1} \ = \ 8.35 \cdot 10^{-5}.
\label{a1a2M10}
\eeq
Substituting these into (\ref{vcnum}) we find that at the end of the
compression phase, i.e. at the time $t=12t_{SC}$, the cloud velocity
is $10\%$ of the postshock velocity $v_{PS}$ and $7.5\%$ of the shock
velocity $v_{S}$. On the other hand, at the end of the re-expansion
phase, i.e. at the time $t=24t_{SC}$ the cloud velocity is $12.66\%$
of $v_{PS}$ and $9.4\%$ of $v_{S}$.

For the case $M_{S} \to \infty$ the above coefficients have the values
\footnote{Note that the assumption here is the same, as in the
discussion of Mach scaling, namely, while increasing the shock Mach
number, we keep the shock front velocity in the stationary reference
frame to be constant.}
\beq
a_{1} \ = \ 1.83 \cdot 10^{-3}; \ a_{2} \ = \ 1.09; \
b_{1} \ = \ 8.51 \cdot 10^{-5}.
\label{a1a2Minf}
\eeq
Substitution into (\ref{vcnum}) gives us the maximum values of the velocity
that a cloud can reach in the case of an infinitely strong shock:
\beq
\begin{array}{llllll}
v_{C,max} & = & 10.1\cdot 10^{-2}v_{PS} & = & 7.55\cdot 10^{-2}v_{S},
& compression \ phase; \\
v_{C,max} & = & 12.8\cdot 10^{-2}v_{PS} & = & 9.57\cdot 10^{-2}v_{S},
& re-expansion \ phase.
\end{array}
\label{vcmax}
\eeq

Similarly, we can determine the values of cloud displacement for the
two cases, considered above. Expression (\ref{l}) for $L_{C}$ can be
rewritten as follows
\beq
L_{C}(t)= \left\{ \begin{array}{ll}
\displaystyle a_{0}c_{1}\Bigg(\frac{t}{t_{SC}}-
\frac{1}{a_{1}} \ ln\Big(\frac{t}{t_{SC}}\Big(\frac{a_{1}}{a_{2}}
\Big)+1\Big)\Bigg),& 0 \le t \le t_{CC} \\
\displaystyle a_{0}c_{1}\Bigg(\frac{t}{t_{SC}}-
c_{2} \ tan^{-1}\Big(\frac{\frac{t}{t_{SC}}-12}{\frac{t}{t_{SC}}c_{4}
+ c_{5}}\Big)-c_{3}\Bigg), & t_{CC} \le t \le t_{CD}
\end{array} \right.,
\label{lcdnum}
\eeq
where for the case $M_{S}=10$ the coefficients $a_{1}$ and $a_{2}$
have the values defined in (\ref{a1a2M10}), and for $M_{S} \to \infty$
the values defined in (\ref{a1a2Minf}).

In the case $M_{S}=10$ the coefficients $c_{1}$, $c_{2}$, $c_{3}$,
$c_{4}$, and $c_{5}$ have the values
\begin{displaymath}
c_{1} \ = \ 1.49; \ c_{2} \ = \ 104.17; \ c_{3} \ = 10.89;
\ c_{4} \ = \ 9.34\cdot 10^{-2}; \ c_{5} \ = \ 113.72.
\end{displaymath}
Substitution into (\ref{lcdnum}) gives us the displacement that the
cloud material undergoes by the end of the compression and
re-expansion phases: $1.6a_{0}$ and $3.5a_{0}$ correspondingly.

In the limiting case $M_{S}\rightarrow \infty$ the values of the
coefficients $c_{1}$, $c_{2}$, $c_{3}$, $c_{4}$, and $c_{5}$ are
the following
\begin{displaymath}
c_{1} \ = \ 1.5; \ c_{2} \ = \ 103.22; \ c_{3} \ = 10.9;
\ c_{4} \ = \ 9.43\cdot 10^{-2}; \ c_{5} \ = \ 112.56.
\end{displaymath}
Substituting these coefficients into (\ref{lcdnum}) we find that by
the end of the compression phase the cloud is displaced by the
distance of $1.65a_{0}$, whereas by the end of the re-expansion phase
the displacement is $3.53a_{0}$.

It is clear from the results, obtained above, that both the velocity and
cloud displacement values in the case $M_{S}=10$ are practically
identical to the maximum values, achieved in the limiting case of $M_{S}
\to \infty$. Therefore, our results obtained for the case
of a Mach 10 shock can be considered as the limiting ones for the
cases of strong shocks.

These results, derived for single clouds or systems with large
separation, are in good agreement with numerical experiments.
Typically the maximum difference between numerical and analytical
values of cloud velocity and position never exceeds $10\%$.  The
analytical results are usually an overestimate of the numerical
ones. This is due to a slight overestimation of the initial
velocity gain after the contact with the external shock front and
because we assumed the cloud cross-section to be constant during
the compression phase, whereas it undergoes a small decrease in
the experiments.

Therefore, the maximum distance a cloud can travel before its
destruction after the initial interaction with a strong shock is
\beq
L_{CD,max} \approx 3.5a_{0}.
\label{lcdmax}
\eeq

\subsubsection{Critical Cloud Separation}

We first define the average cloud separation, projected on to the
direction of the flow $\langle\Delta x_{N} \rangle$ and
perpendicular to it $\langle\Delta y_{N} \rangle$, for a given
cloud distribution, \beq \langle\Delta x_{N} \rangle =
\frac{\displaystyle\sum_{i,j>i}^{N} |x_{i}-x_{j}|}
{C_{2}^{N}}=\frac{2}{(N-1)N}\displaystyle\sum_{i,j>i}^{N}|x_{i}-x_{j}|,
\label{xav} \eeq \beq \langle\Delta y_{N} \rangle =
\frac{\displaystyle\sum_{i,j>i}^{N} |x_{i}-x_{j}|}
{C_{2}^{N}}=\frac{2}{(N-1)N}\displaystyle\sum_{i,j>i}^{N}|y_{i}-y_{j}|.
\label{yav} \eeq We can also define a maximum cloud separation
projected on to the direction of the flow, or the ``cloud layer
thickness'', \beq (\Delta x_{N})_{max} =
\displaystyle\max_{i,j\in\{1,N\}}\{|x_{i}-x_{j}|\}. \label{xmax}
\eeq

Now we are in a position to give a precise definition of the
``thin-layer'' and ``thick-layer'' systems. We define a
distribution of clouds to belong to the subset $\Lambda_{I}$ if
its maximum cloud separation $(\Delta x_{N})_{max}$ does not
exceed the cloud destruction length $L_{CD}$.  The distribution
belongs to the subset $\Lambda_{M}$ in all other cases: \beq
\begin{array}{c}
\Lambda_{I} \equiv\{\lambda_{N} : (\Delta x_{N})_{max} \le L_{CD}\}, \\
\Lambda_{M} \equiv\{\lambda_{N} : (\Delta x_{N})_{max} > L_{CD}\}.
\end{array}
\label{2classes}
\eeq

A more intuitive way to look at this classification is the
following. As we have seen, a cloud interacting with the postshock
flow re-expands and breaks up before it proceeds into the mixing
phase. The above criterion tells us if any cloud or a row of
clouds will complete its destruction phase prior to encountering
any other clouds located downstream. The definition
(\ref{2classes}) appears to draw rather accurately the line
between cloud systems of two types.

In practice the maximum cloud separation $(\Delta x_{N})_{max}$
(eq. \ref{xmax}) is simply the thickness of the layer of
inhomogeneities in a real system and should be compared against
the cloud destruction length. This thickness can be obtained from
the observations of a particular object or it can be found
analytically, e.g. via consideration of instabilities at the
interface between two flows.

Having defined the two classes, or subsets, of cloud distributions
we now consider the behaviour of the clouds in each class. First
we consider $\Lambda_{I}$, the ``single-row'' distributions. On
average, by the time the clouds are displaced by the distance
$L_{CD}$, all of them will be destroyed and will proceed to the
mixing phase. Thus time of the destruction should be approximately
$t_{CD} \simeq t_{SD}$.

The question arises whether clouds will interact during the process of
re-expansion and destruction. We can give a formal criterion for
this. Consider two clouds with separation $\langle\Delta y_{N} \rangle
\ = \ d$ and $(\Delta x_{N})_{max} \leq L_{CD}$. Both clouds will expand 
laterally at the velocity $v_{exp}$ defined in (\ref{cc}).
Consequently the time for the clouds to come into contact is
\beq
t_{merge} \approx \frac{d-2a_{0}}{2v_{exp}}.
\eeq 
Such re-expansion starts after the cloud compression phase, i.e. after
the time $t_{CC}$ and cannot proceed beyond the cloud destruction time
$t_{CD}$. Therefore, setting $t_{merge}=t_{CD}-t_{CC}$ we find the
following critical cloud separation transverse to the global flow
\beq
d_{crit}=2\big(a_{0}+v_{exp}(t_{CD}-t_{CC})\big).
\label{dcrit1}
\eeq
Substituting (\ref{cc}) explicitly for the expansion velocity and
(\ref{tcd}) for the cloud destruction time into (\ref{dcrit1}) we
obtain
\beq
d_{crit}=2a_{0}\Bigg\{\frac{t_{CD}-t_{CC}}{t_{SC}}
\Bigg(\frac{F_{c1}F_{st}}{\chi}\Bigg)^{\frac{1}{2}}\Bigg(\frac{3\gamma
(\gamma - 1)}{\gamma + 1}\Bigg)^{\frac{1}{2}}+1\Bigg\}.
\label{dcrit}
\eeq
In other words clouds whose separation transverse to the flow is less
than $d_{crit}$ will come into contact and merge before their
destruction is completed.  Therefore, their evolution during the
destruction phase (and for the most part of the re-expansion phase)
can not be considered as the evolution of two independent clouds.

The critical separation does not depend on the global shock Mach
number in consistency with the Mach scaling, discussed above.
Therefore, this parameter is universal for all strong shocks and for
all possible distributions from the subset $\Lambda_{I}$. For the case
$\gamma = 5/3$ and $\chi = 500$ we find the critical cloud separation
to be approximately
\beq
d_{crit}\approx 4.2a_{0}.
\label{dcritnum}
\eeq

For cloud distributions from the subset $\Lambda_{I}$ which have an
average separation $\langle\Delta y_{N} \rangle \gg d_{crit}$, the
evolution of the system will proceed in the \emph{non-interacting
regime}. On the other hand, for the distributions, for which
$\langle\Delta y_{N} \rangle \lesssim d_{crit}$, the cloud-cloud
interactions are important throughout the re-expansion and destruction
phases placing them in an \emph{interacting regime}.

\begin{figure}
\epsscale{0.65}
\plotone{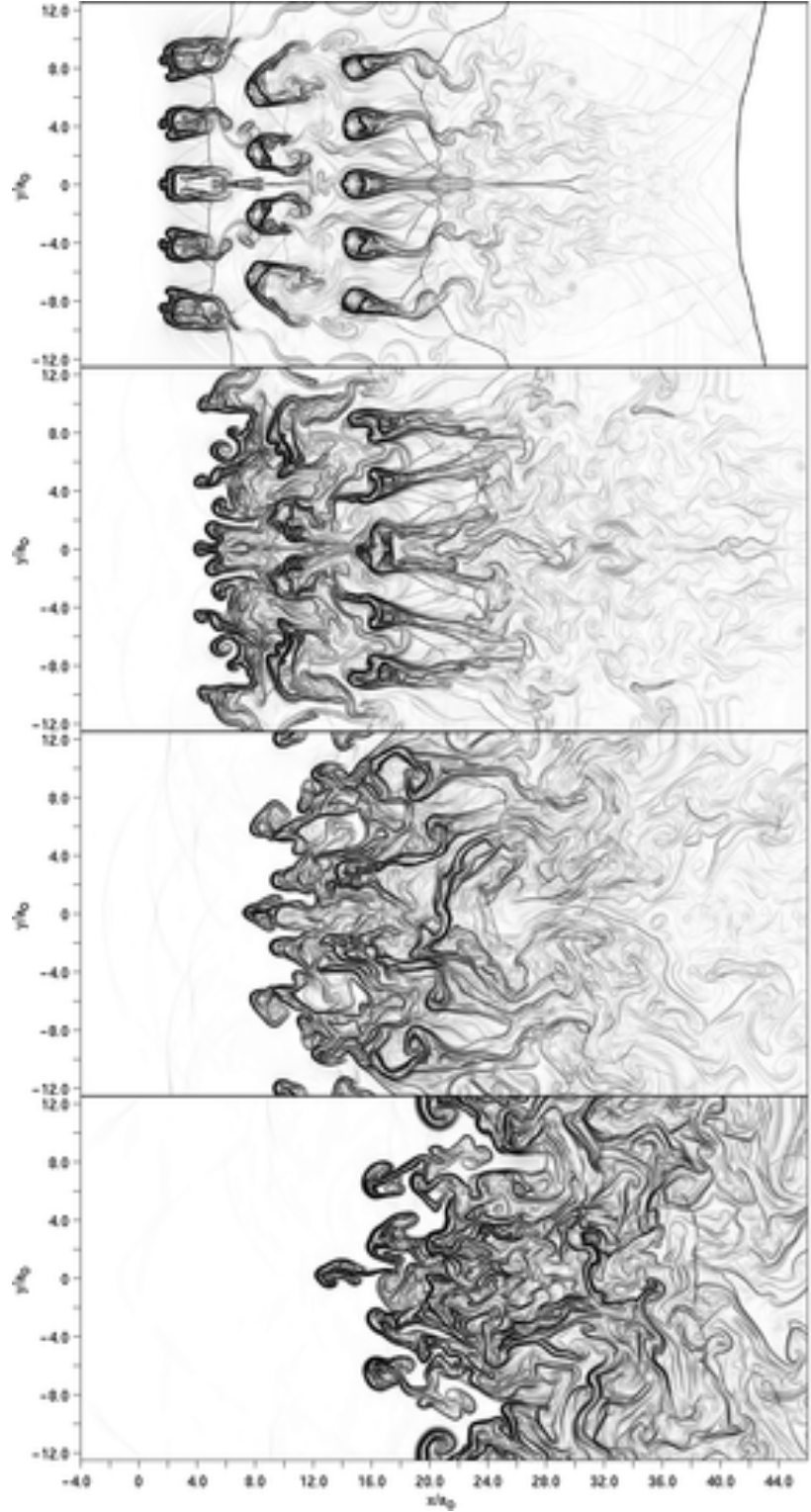}
\caption{Run $M14$. Time evolution of a system, containing
fourteen identical clouds in a regular distribution and interacting
with a $M_{S}=10$ shock wave. Shown are the synthetic Schlieren images
of the system at times $22.47 \ t_{SC}$, $35.23 \ t_{SC}$, $50.54 \
t_{SC}$, $69.09 \ t_{SC}$.
\label{14clump}}
\end{figure}

\begin{figure}
\epsscale{0.65}
\plotone{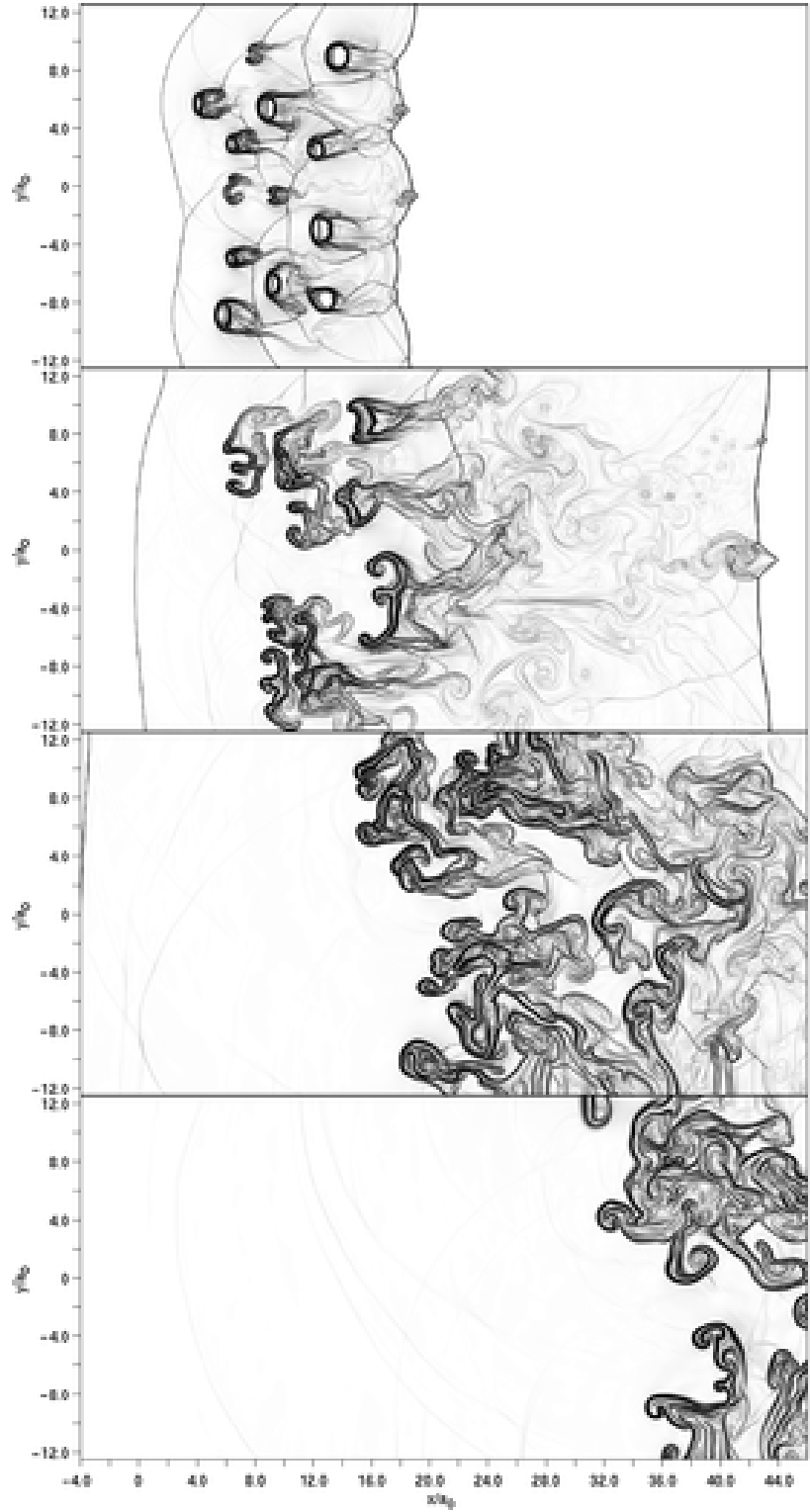}
\caption{Run $M14_{r}$. Time evolution of a system, containing
fourteen clouds in a random distribution and interacting with a
$M_{S}=10$ shock wave. Shown are the synthetic Schlieren images of the
system at times $9.71 \ t_{SC}$, $22.47 \ t_{SC}$, $45.43 \ t_{SC}$,
$69.09 \ t_{SC}$.
\label{14random}}
\end{figure}

\begin{figure}
\plotone{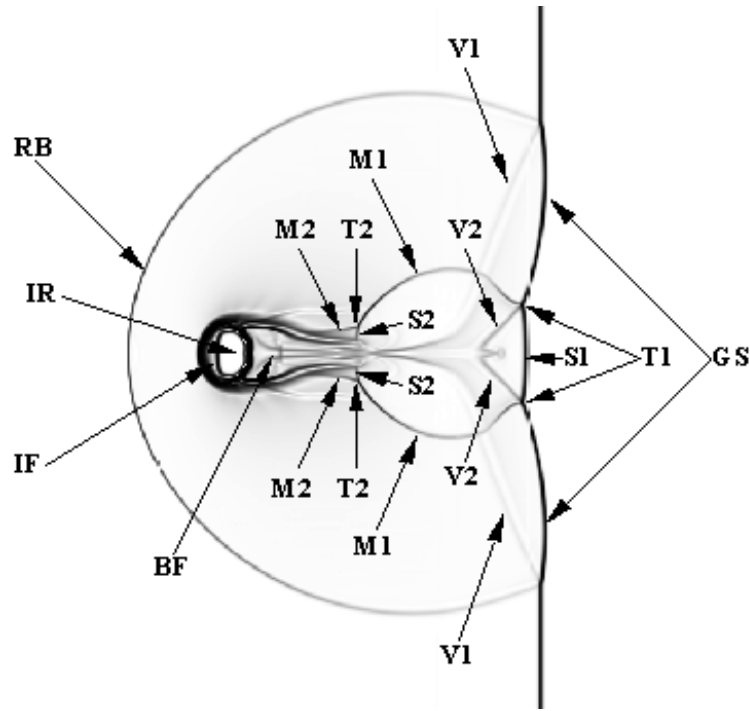}
\caption{Flow structure during the initial compression phase. Shown is the
Schlieren image of the run $M1$ at time $5.1 \ t_{SC}$.
$GS$ - external global forward shock;
$RB$ - external reverse bow shock;
$IF$ - internal forward shock;
$IR$ - internal reverse shock;
$BF$ - back flow, caused by global forward shock convergence on the symmetry axis;
$V1$ - primary vortex sheets, caused by regular reflection of the bow shock;
$M1$ - primary Mach reflected shocks, caused by Mach reflection of the global
forward shock at the symmetry axis;
$S1$ - primary Mach stem (more precisely, two primary Mach stems);
$T1$ - primary triple points;
$V2$ - secondary vortex sheets, caused by the primary Mach reflection
of the global forward shock (note the two stem bulges formed at the base
of the secondary vortex sheets near the symmetry axis);
$M2$ - secondary Mach reflected shocks;
$S2$ - secondary Mach stems;
$T2$ - secondary triple points.
\label{shocks}}
\end{figure}

\begin{figure}
\epsscale{0.65}
\plotone{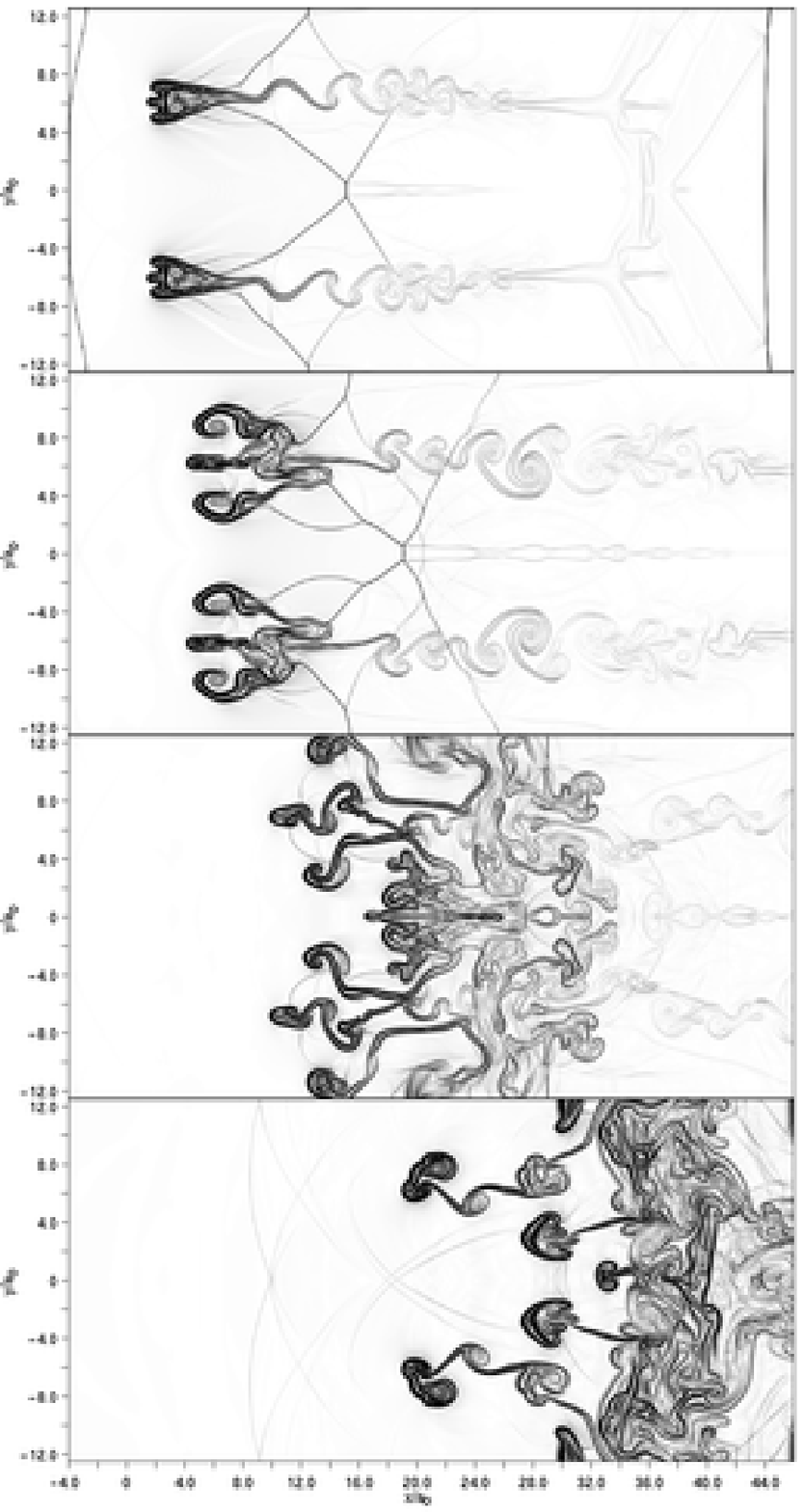}
\caption{Run $A2$. Illustration of the \emph{non-interacting regime}
of cloud evolution: interaction of a $M_{S}=10$ shock wave with a
system of two identical clouds with the cloud center separation of
$12.0 \ a_{0} \approx 2.86 \ d_{crit}$. Shown are the synthetic Schlieren
images of the system at times $22.47 \ t_{SC}$, $35.23 \ t_{SC}$,
$50.54 \ t_{SC}$, $68.40 \ t_{SC}$.
\label{2clumpwide}}
\end{figure}

\begin{figure}
\epsscale{0.65}
\plotone{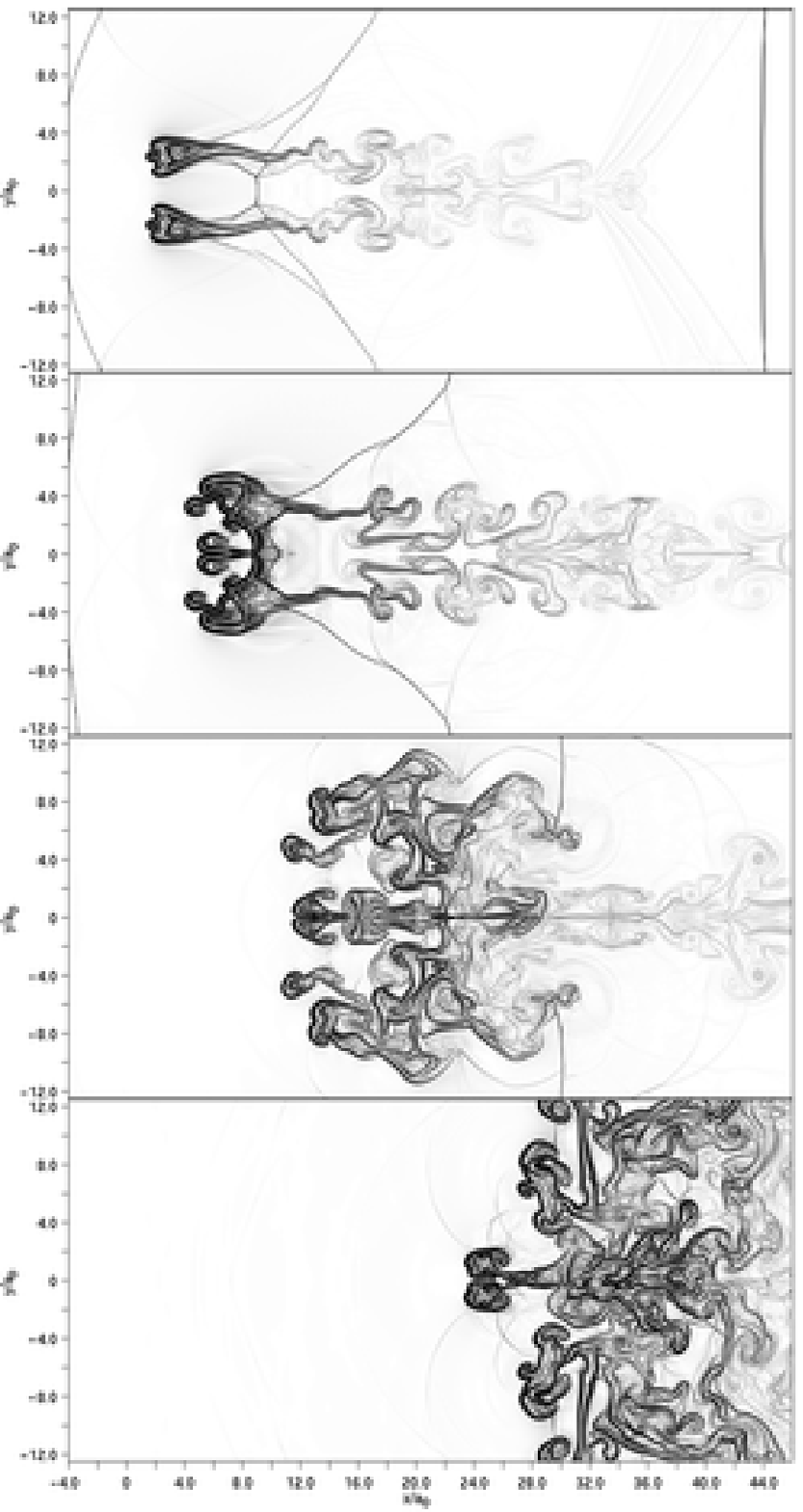}
\caption{Run $M2$. Illustration of the \emph{interacting regime} of
cloud evolution: interaction of a $M_{S}=10$ shock wave with a system
of two identical clouds with the cloud center separation of $4.0 \ a_{0}
\approx 0.95 \ d_{crit}$. Shown are the synthetic Schlieren images of the
system at times $22.47 \ t_{SC}$, $35.23 \ t_{SC}$, $50.54 \ t_{SC}$,
$68.40 \ t_{SC}$.
\label{2clump}}
\end{figure}

\begin{figure}
\epsscale{1.00}
\plotone{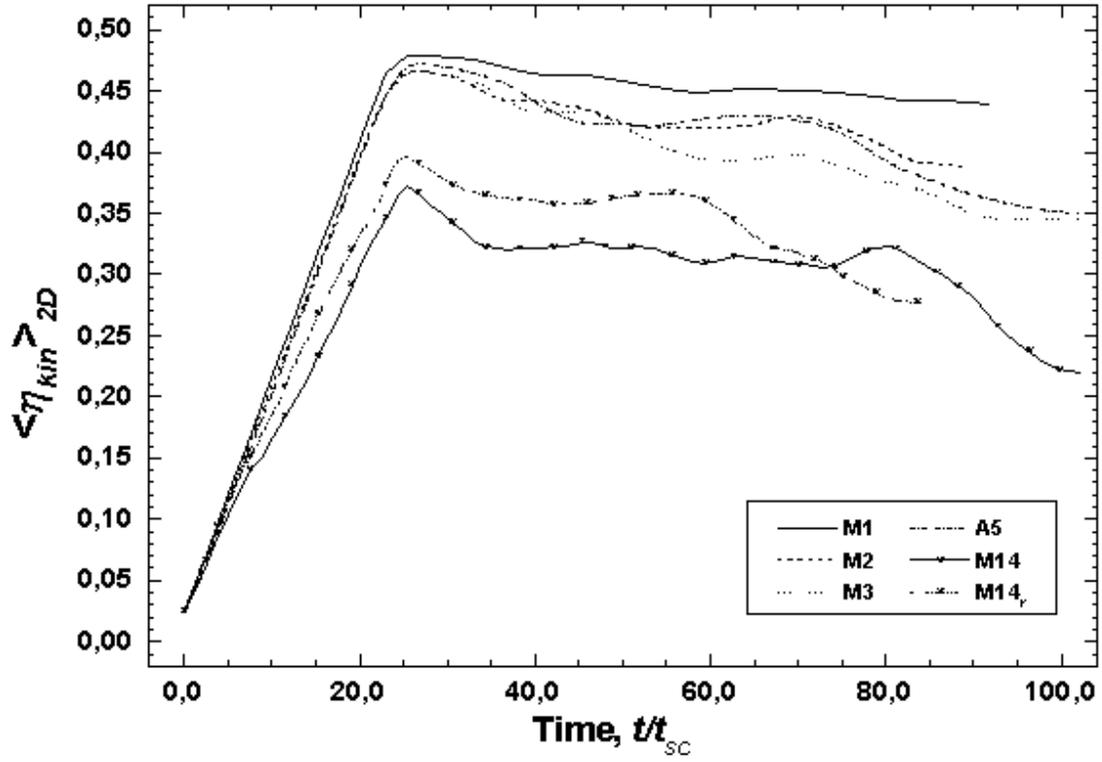}
\caption{Time evolution of the global average of the kinetic energy fraction
$\langle\eta_{kin}\rangle_{2D}$ for the runs $M1$, $M2$, $M3$, $A5$,
$M14$, $M14_{r}$.
\label{ekin}}
\end{figure}

\begin{figure}
\epsscale{1.00}
\plotone{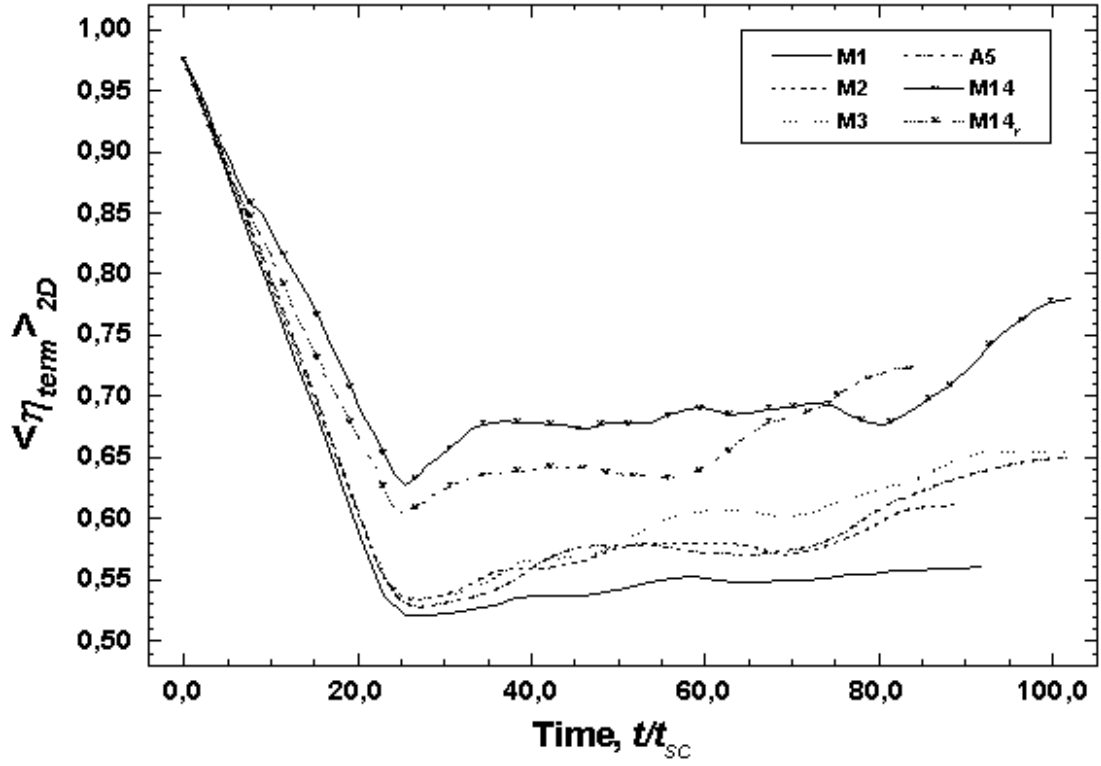}
\caption{Time evolution of the global average of the thermal energy fraction
$\langle\eta_{term}\rangle_{2D}$ for the runs $M1$, $M2$, $M3$, $A5$,
$M14$, $M14_{r}$.
\label{eterm}}
\end{figure}

\begin{figure}
\epsscale{1.00}
\plotone{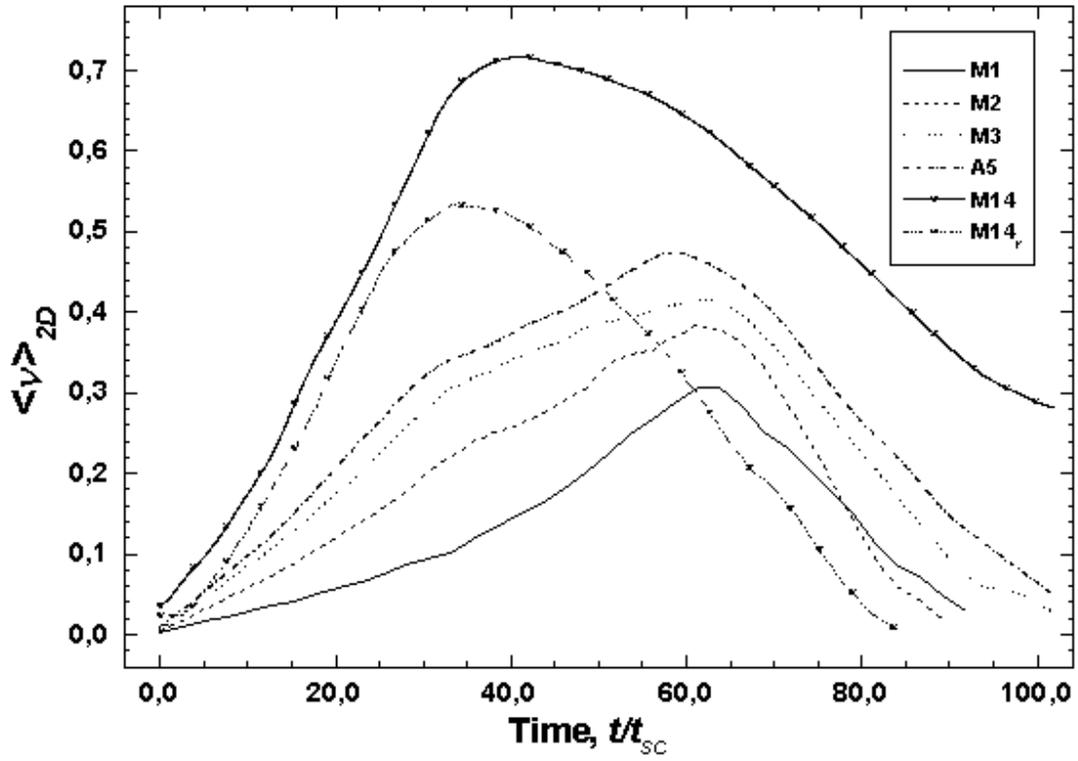}
\caption{Time evolution of the global average of the volume filling factor
$\langle\nu\rangle_{2D}$ for the runs $M1$, $M2$, $M3$, $A5$, $M14$,
$M14_{r}$.
\label{fill}}
\end{figure}

It is more difficult to formulate a unified criterion for the behavior
of the systems in the class $\Lambda_{M}$. When $\langle\Delta x_{N}
\rangle > L_{CD}$ such systems can be considered as a set of thin
layers with an average separation $L_{CD}$, i.e. each row can be
considered as a system from subset $\Lambda_{I}$. Consider, for
example, the run $M_{14}$, presented in Figure~\ref{14clump}. From
Table (1), the average separation $\langle\Delta x_{N}\rangle$ for
this run is equal to 7, i.e. $\langle\Delta x_{N}\rangle >
L_{CD}$. Indeed, the evolution of the leftmost row of clouds proceeds
as a simple single row case, and its destruction is completed by the
time $t_{CD}$. This results in the fractional kinetic energy reaching
a maximum at time $t_{CD} \approx 24t_{SC}$ (see
Figure~\ref{ekin}). However, it is clear from Figure~\ref{14clump}
that the evolution of the downstream rows is altered by the
destruction of the leftmost one. Therefore, when $\langle\Delta
x_{N}\rangle > L_{CD}$ one must account for the fact that the
destruction of an upstream layer of clouds will change the properties
of the global flow for the next, downstream layer.  The new averaged
values of the velocity, density, and pressure in the global flow
should then be used as an input for the analysis of the downstream
cloud layer.

\subsection{Mass loading}

One of the principal questions concerning the effects of
shock/cloud-system interactions is the role of mass-loading
\citep{HD88}. Mass-loading is defined as the feeding of material
into the global flow by nearly stationary clouds. Analytical
studies have predicted a number of important changes when
mass-loading occurs.  The most important of these is the
transition of the flow to a transonic regime \citep{HD86},
\citep{HD88}, \citep{DH92}, \citep{DH94}. In our numerical
experiments we consider if mass-loading indeed is prominent.

Mass loading can occur only from time $t=0$ up to the moment of cloud
destruction at time $t=t_{CD}$. In our experiments the cloud
destruction time is fairly short compared with the total age of most
relevant astrophysical objects. Indeed, cloud destruction is
practically completed by the time the shock wave reaches the right
boundary of the computational domain, i.e. by the time the shock wave
travels the distance of about 20-30 cloud sizes. This could, for
example, be compared with clump systems in planetary nebulae.
Assuming typical size for PNe clouds to be about 100 a.u. (which is
the size of cometary knots in NGC 7293 \citep{ODell98}), a density
contrast 500, and a shock wave velocity 100 \kms, we find that clouds
get completely destroyed within approximately $100-150$ years. This is
much less than the typical age of the planetary nebulae ($10^{4} \ - \
10^{5}$ yrs.).

Thus clouds with low density contrast $\chi_{i} \approx 10 - 100$ can
not provide significant mass-loading due to the ease in which they are
advected and destroyed by the global flow. On the other hand, clouds
with higher density contrasts $\chi_{i} > 100$ retain their low
velocities with respect to the global flow for much longer periods of
time and, therefore, may potentially be efficient mass-loading
sources. However, it should be noted, that this higher relative
velocity of a cloud increases the efficiency of the instability
formation, thereby promoting cloud destruction and its mixing with the
flow.

\begin{figure}
\epsscale{0.50}
\plotone{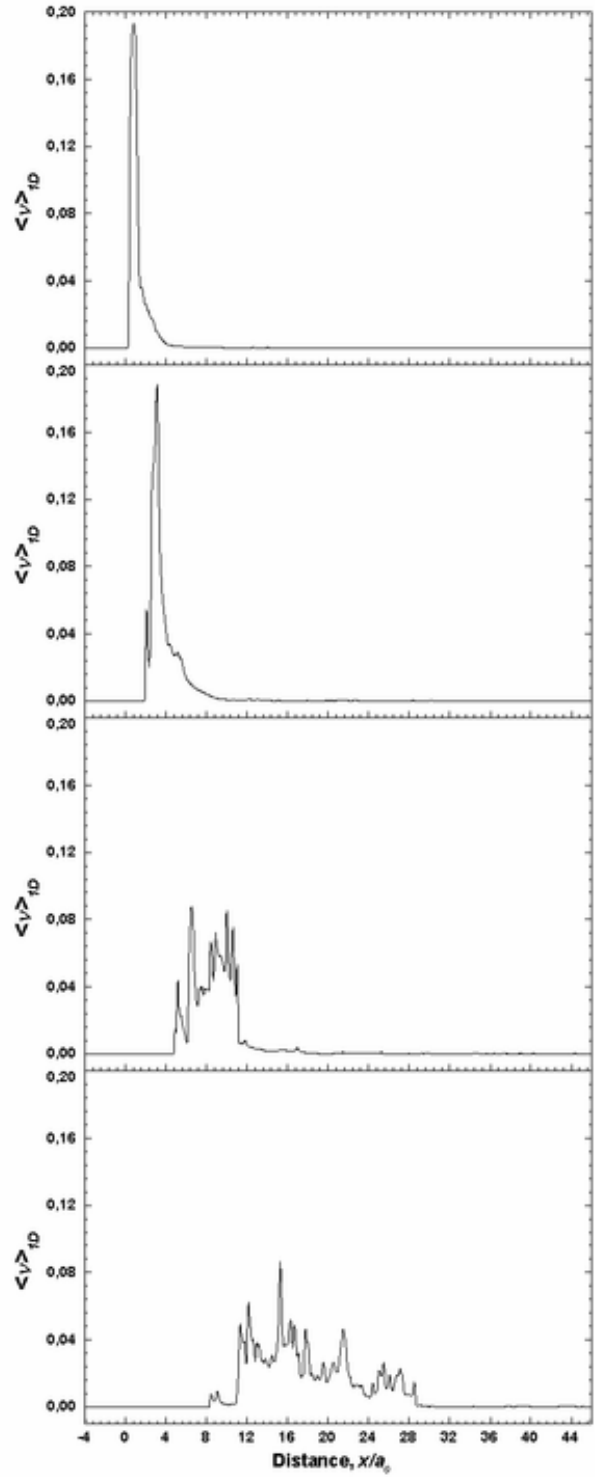}
\caption{Distribution of cloud material along the horizontal dimension of the
computational domain for the run $M3$. Shown are the one-dimensional spatial
averages of the volume filling factor $\langle\nu\rangle_{1D}$ at times
$12.26 \ t_{SC}$, $25.02 \ t_{SC}$, $37.78 \ t_{SC}$, $50.54 \ t_{SC}$.
\label{massload}}
\end{figure}

We can also consider the amount of mass seeded into the flow, i.e.
stripped of from the clouds and assimilated into the global flow,
before cloud destruction. Typically, in our experiments the amount of
seeded cloud material does not exceed a few percent of the total cloud
mass, which is unlikely to be enough to switch the flow into a
mass-loaded regime. Figure~\ref{massload} shows the distribution of
cloud material along the direction of the flow or, to be more precise,
the distribution of the parameter $\langle\nu\rangle_{1D}(x)$ for the
three cloud run $M_{3}$. There the clouds have the separation
$\langle\Delta y_{N} \rangle \approx 0.95d_{crit}$. The first graph
corresponds to the end of the compression phase, while the second
corresponds to the end of the destruction phase. The graphs show that
cloud material remains localized in the vicinity of the cloud cores
until the moment of cloud destruction and the system does not exhibit
any significant mass-loading. Moreover, the graphs 3 and 4 of
Figure~\ref{massload}, showing cloud material distribution early in
the mixing phase, indicate that even after destruction cloud material
remains localized within the region of about 8 cloud radii and retains
almost the same average velocity with respect to the global flow. Only
further on in the mixing phase does cloud material spread
significantly.

Concluding, we may say that for the cloud density contrast values in
the range $\chi_{i} \approx 10 - 1000$ and practically all values of
the global shock wave Mach number, the flows are not likely to be
subject to mass loading. These flows will be dominated by the mixing
of cloud material with the global flow that occurs after cloud
destruction. Systems with very dense clouds $\chi_i >> 1000$ may
provide sites suitable for mass loading.  Future numerical studies
should be able to confirm this.

\section{CONCLUSIONS}

We have numerically investigated the interaction of a strong,
planar shock wave with a system of inhomogeneities. These "clumps"
are considered to be infinitely long cylinders embedded in a
tenuous, cold ambient medium. We have assumed constant conditions
in the global postshock flow, thereby constraining the maximum
size of the clouds only by the condition of the shock front
planarity. Our results are applicable to strong global shocks with
Mach numbers $3 \lesssim M_{S} \lesssim 1000$. The range of the
applicable cloud/ambient density contrast values is $10 - 1000$.

We considered four major phases of the cloud evolution due to the
interaction of the global shock and postshock flow with a system of
clouds.  These are: initial compression phase, re-expansion phase,
destruction phase, and mixing phase. We describe a simple model for
the cloud acceleration during the first three phases, i.e. prior to
its destruction, and derive expressions for the cloud velocity and
displacement. The results of that model are in excellent agreement
with the numerical experiments. The difference in the values of cloud
velocity and displacement between analytical and numerical results is
$\lesssim 10\%$. The maximum cloud displacement due to its interaction
with a strong shock (prior to its destruction) does not exceed $3.5$
initial maximum cloud radii. The maximum cloud velocity is not more
than $10\%$ of the global shock velocity.

The principal conclusion of the present work is that the set
$\Lambda$ of all possible cloud distributions can be subdivided
into two large subsets $\Lambda_{I}$ and $\Lambda_{M}$. The first
subset is $\Lambda_{I}$, thin-layer systems.  This subset is
defined by the condition that the maximum cloud separation along
the direction of flow, or the cloud layer thickness, is not
greater than the cloud destruction length $(\Delta x_{N})_{max}
\le L_{CD}$. The  thick-layer systems $\Lambda_{M}$, are defined
by the condition $(\Delta x_{N})_{max} > L_{CD}$. The evolution of
cloud distributions within each subset exhibit striking similarity in
behaviour. We conclude that the evolution of a system of clouds
interacting with a strong shock depends primarily on the total
thickness of the cloud layer and the cloud distribution in it, as
opposed to the total number of clouds or the total cloud mass present
in the system. The key parameters determining the type of the cloud
system evolution are therefore the critical cloud separation
transverse to the flow $d_{crit}$ (this is also the critical linear
cloud density in the layer), and the cloud destruction length
$L_{CD}$.

For a given astrophysical situation our results indicate that one
might determine, either from observations or from theoretical
analysis, the thickness of the cloud layer $(\Delta x_{N})_{max}$.
This will then determine the class of the given cloud
distribution, $\Lambda_{I}$ or $\Lambda_{M}$. For cloud
distributions from the set $\Lambda_{I}$ with average cloud
separation $\langle \Delta y_{N} \rangle > d_{crit}$ evolution of
the clouds during the compression, re-expansion, and destruction
phases will proceed in the noninteracting regime and the formalism
for a single cloud interaction with a shock wave (e.g. KMC,
\citep{Jones96}, \citep{MacLow94}, \citep{Lim99}) can be used to 
describe the system. On the other hand, if the cloud separation is
less than the critical distance, the clouds in the layer will merge
into a single structure before their destruction is completed. Though
throughout the compression phase they can still be considered
independently of each other, their evolution during the re-expansion
and destruction phases clearly proceeds in the interacting regime.

When the distribution belongs to the subset $\Lambda_{M}$ it is
necessary to determine the average cloud separation projected onto
the direction of the flow $\langle \Delta x_{N} \rangle$, defined
by (\ref{xav}) above, and compare it against $L_{CD}$: if $\langle
\Delta x_{N} \rangle > L_{CD}$ evolution of the cloud system can
be roughly approximated as the evolution of a set of distributions
from the subset $\Lambda_{I}$ and the above ``thin-layer case''
analysis applies. If, on the other hand, $\langle \Delta x_{N}
\rangle \le L_{CD}$ (especially if $\langle \Delta y_{N} \rangle <
d_{crit}$) the system evolution is dominated by cloud interactions
and a thin layer formalism is inappropriate.

Finally we have considered the role of mass-loading. Here our
principal conclusion is that the mass-loading is not significant
in the cases of strong shocks interacting with a system of
inhomogeneities for density contrasts in the range $10-1000$. In
part this is due to short survival times of clouds under such
conditions, and in part due to the very low mass loss rates of the
clouds even during the times prior to their destruction. Mass loading
may well be important in higher density clouds \citep{DH94}.

The major limitation of our current work is the purely hydrodynamic
nature of our analysis that does not include any consideration of
magnetic fields. As it was discussed in section 3.1.4, cold dense
inhomogeneities (clouds) embedded in tenuous hotter medium are
inherently unstable against the dissipative action of diffusion and
thermal conduction. This evaporates the clouds on the timescales
comparable to, or shorter than, the timescales of the dynamical
evolution of the system. It was suggested that the magnetic fields may
play a stabilizing role against the action of the dissipative
mechanisms. Although weak magnetic fields, that are dynamically
insignificant up to the moment of cloud destruction, can inhibit
thermal conduction and diffusion, those magnetic fields may become
dynamically important due to turbulent amplification during the mixing
phase. A fully magnetohydrodynamic description of the interaction of a
strong shock with a system of clouds will need to be carried forward
in future works.

\acknowledgements

This work was supported in part by the NSF grant AST-9702484 and
the Laboratory for Laser Energetics under DOE sponsorship.

The most recent results and animations of the numerical experiments,
described above and not mentioned in the current paper, can be found
at \url{www.pas.rochester.edu$/^{\sim}$wma}.


\clearpage

\begin{deluxetable}{cccccc}
\tablecaption{Summary of the Runs Discussed \label{Runs}}
\tabletypesize{\small}
\tablewidth{0pt}
\tablehead{
\colhead{Run} & \colhead{$\#$ of clouds \tablenotemark{a}} &
\colhead{Distribution}
 & \colhead{$\#$ of rows} & \colhead{x-spacing \tablenotemark{b}} &
\colhead{y-spacing \tablenotemark{c}}}
\startdata
M1 & 1 & regular & 1 & - & - \\ M2 & 2 & regular & 1 & - & 4 \\ A2 & 2
& regular & 1 & - & 12 \\ M3 & 3 & regular & 1 & - & 4 \\ A5 & 5
& regular & 1 & - & 4 \\ M14 & 14 & regular & 3 & 7 & 4 \\ $M14_{r}$ &
14 & random & 3 & 3.5\tablenotemark{d} & 3.5\tablenotemark{d} \\
\enddata
\tablenotetext{a}{ Total number of clouds present in the system.}
\tablenotetext{b}{ Spacing between the centers of clouds in \emph{two different
rows}, projected onto the x-axis, in the units of the maximum cloud
radius $a_{max}$}
\tablenotetext{c}{ Spacing between the centers of clouds in \emph{the same row},
projected onto the y-axis, in the units of the maximum cloud radius $a_{max}$}
\tablenotetext{d}{ Maximum absolute spacing between the cloud centers in
\emph{any} direction.}
\end{deluxetable}


\clearpage

\clearpage

\clearpage

\clearpage

\clearpage

\clearpage

\clearpage

\clearpage

\clearpage

\clearpage

\clearpage

\clearpage

\end{document}